\documentclass[twocolumn,amsmath,amssymb,floatfix,pre,floatfix,unsortedaddress]{revtex4-1} 

\usepackage{graphicx,color}
\definecolor{brown}{rgb}{0.63,0.27,0.18}
\definecolor{orange}{rgb}{1.00,0.65,0.00}

\usepackage{moresize}
\usepackage{dcolumn}
\usepackage{bm,multirow}

\makeatletter
\newcommand*{\balancecolsandclearpage}{%
  \close@column@grid
  \twocolumngrid
}
\makeatother

\newcommand{\be}{\begin{equation}}
\newcommand{\ee}{\end{equation}}
\usepackage{color}

\begin{document}

\newcommand {\rsq}[1]{\left< R^2 (#1)\right.}
\newcommand {\rsqL}{\left< R^2 (L) \right>}
\newcommand {\rsqbp}{\left< R^2 (N_{bp}) \right>}
\newcommand {\Nbp}{N_{bp}}
\newcommand {\etal}{{\em et al.}}
\newcommand{\Ham}{{\cal H}}
\newcommand{\AngeloComment}[1]{\textcolor{red}{(AR) #1}}
\newcommand{\AndreaComment}[1]{\textcolor{blue}{A: #1}}
\newcommand{\JanComment}[1]{\textcolor{green}{(JS) #1}}

\newcommand{\NewText}[1]{\textcolor{orange}{#1}}
\newcommand{\scs}{\ssmall}

\newcommand{\Tau}{\mathrm{T}}



\title{Nanorheology of active-passive polymer mixtures is topology-sensitive}

\author{Andrea Papale}
\email{andrea.papale@ens.psl.eu}
\altaffiliation{Current affiliation: Group of Data Modeling and Computational Biology, IBENS-PSL \'Ecole Normale Sup\'erieure, Paris, France.}
\affiliation{SISSA - Scuola Internazionale Superiore di Studi Avanzati, Via Bonomea 265, 34136 Trieste, Italy}

\author{Jan Smrek}
\email{jan.smrek@univie.ac.at}
\affiliation{Faculty of Physics, University of Vienna, Boltzmanngasse 5, A-1090 Vienna, Austria}

\author{Angelo Rosa}
\email{anrosa@sissa.it}
\affiliation{SISSA - Scuola Internazionale Superiore di Studi Avanzati, Via Bonomea 265, 34136 Trieste, Italy}

\date{\today}

\begin{abstract}
We study the motion of dispersed nanoprobes in entangled active-passive polymer mixtures.
By comparing the two architectures of linear {\it vs.} unconcatenated and unknotted circular polymers,
we demonstrate that novel, rich physics emerge.
For both polymer architectures, nanoprobes of size smaller than the entanglement threshold of the solution move faster as activity is increased and more energy is pumped in the system.
For larger nanoprobes, a surprising phenomenon occurs: while in linear solutions they move qualitatively as before, in active-passive ring solutions nanoprobes {\it decelerate} with respect to the purely passive conditions.
We rationalize this effect in terms of the non-equilibrium, topology-dependent association (clustering) of nanoprobes to the cold component of the ring mixture
reminiscent of the recently discovered [Weber {\it et al.}, Phys. Rev. Lett. {\bf 116}, 058301 (2016)] phase separation in scalar active-passive mixtures.
We conclude with a potential connection to the microrheology of the chromatin in the nuclei of the cells.
\end{abstract}

\maketitle

{\it Introduction} --
In recent years, micro- and nanorheology have emerged as promising tools to probe, non-invasively, the viscoelastic properties of complex colloidal systems, polymer materials and polymer solutions~\cite{MasonWeitz1995,PuertasJPCM2014}, even the interior of the cells~\cite{HameedShivashankarPlosOne2012}, through monitoring the time dependence of the mean-square displacement of tagged nanoprobes.
In fact, as confirmed by recent numerical work as well as theoretical considerations~\cite{KalathiGrestPRL2014,NahaliRosa2016,NahaliRosa2016,GeRubinsteinMacromolecules2017,RabinGrosbergNanoRheol2019}, the motion of nanoprobes, especially in polymer solutions, results from the interplay between the chemo-physical properties of the chains (density and flexibility) and the ``unavoidable'' {\it topological constraints} (TC's).
Rooted in the mutual uncrossability between nearby chains TC's (a.k.a. entanglements) force polymers to slide past each other and determine~\cite{DeGennes1979,DoiEdwardsBook,RubinsteinColbyBook,WangPolymersReview2017} the characteristic slow viscoelastic relaxation of the compound.

Recently, a lot of effort has been dedicated to investigate systems of so called {\it active polymers}~\cite{WinklerGompper2017}, namely polymers which are maintained in a stationary, out-of-equilibrium state owing to the constant influx of some form of external energy into the system~\cite{WinklerGompper2017,SmrekKremerPRL2017,SmrekEntropy2018,Locatelli-PRL2018,FoglinoSM2019,IlkerJoannyPRR2020,ChubakNatComm2020,Locatelli-PRL2021}.
As in more general active systems~\cite{SolonCatesTailleurNatPhys2015,BechingerRMP2016}, in active polymers non-trivial physical phenomena arise from activity-induced shift to otherwise inaccessible ``corners" of the phase space, hence the applicability of traditional notions from equilibrium thermodynamics is neither expected nor it proves to be adequate~\cite{SolonCatesTailleurNatPhys2015}.

Biological polymers are probably the best examples in this regard:
for instance,
the protein-DNA chromatin fiber in the nucleus of any living cell is systematically subject to processes like transcription, remodelling, repairing or loop extrusion~\cite{SanbornPNAS2015,GoloborodkoBJ2016,Stam2019} which require free energy consumption and dissipation at the fiber level
and induce stronger-than-thermal velocity fluctuations~\cite{Zidovska_PNAS13,SmrekKremerPRL2017}.
Of course, all these effects act synergistically with all the afore-mentioned polymer features, especially the built-in long lasting TC's which are held~\cite{Grosberg1993,RosaPLOS2008,VettorelPhysToday2009,RosaEveraersPRL2014,GrosbergSoftMatter2014,RepProgPhys_RingsChromosomes_2014} responsible for spontaneous chromatin segregation into loopy, compact ({\it i.e.}, ``territorial''~\cite{CremerBrosReview2001}) conformations.

Despite being a highly promising tool to probe properties of emerging active polymeric materials, no systematic study of nanoprobe motion in this context has been attempted.

Motivated by these considerations, in this work we employ extensive molecular dynamics computer simulations of the {\it two}-diffusivities dynamic particle model first introduced in Refs.~\cite{weberweberfreyPRL2016,SmrekKremerPRL2017} and monitor the kinetic properties of nanoprobes dispersed in polymer solutions at {\it high polymer concentrations} and in {\it non-equilibrium conditions}.
In particular, we concentrate on three main aspects of the problem which, at present, remain completely unexplored:
(i) the role of nanoprobe size,
(ii) the role of monomer diffusivities
and
(iii) most importantly, the role of chain topology.
With regard to the latter, for (a) their ``historical'' relevance~\cite{DeGennes1979,DoiEdwardsBook,RubinsteinColbyBook,WangPolymersReview2017} as well as (b) their connection~\cite{Grosberg1993,RosaPLOS2008,VettorelPhysToday2009,RosaEveraersPRL2014,GrosbergSoftMatter2014,RepProgPhys_RingsChromosomes_2014} to the physics of DNA inside the cells, we restrict our discussion here to the simplest chain topologies, namely: entangled linear chains and unknotted and unconcatenated ring polymers in concentrated solutions.
Similarly to recent work~\cite{KalathiGrestPRL2014,NahaliRosa2016,GeRubinsteinMacromolecules2017,RabinGrosbergNanoRheol2019} where nanoprobes move is purely passive polymer media, we show that nanoprobe diffusivity depends in quite a non-trivial way on chain architecture, but in the present non-equilibrium situation the physics is way more subtle and richer:
in particular, probing the active-passive polymer mixtures with nanoprobes of different sizes can not only inform on the polymer topology, but also on the level of activity, local architecture and demixing tendencies.

{\it Model and methods} --
Polymers in solution are modeled according to the generalized Kremer-Grest-like~\cite{KremerGrestJCP1990} bead-spring polymer model considered in previous works~\cite{RosaPLOS2008,RosaBJ2010,RosaEveraersPRL2014},
and the solutions are accompanied by nanoprobes of different diameters.
Nanoprobe-nanoprobe and nanoprobe-polymer purely repulsive interactions are modeled by the phenomenological expressions introduced by Everaers and Ejtehadi~\cite{EveraersEjtehadi2003} and employed in previous works~\cite{ValetRosa2014,NahaliRosa2016,NahaliRosa2018,PapaleRosaPhysBiol2019}.
A more complete account of these potentials is presented in the Supplemental Material (SM).

We have simulated polymer solutions consisting of $M=80$ linear chains or rings, where each polymer is made of $N=500$ monomers of diameter $\sigma$.
Each solution is complemented by the additional presence of $N_{\rm np}=100$ nanoprobes of variable diameters $d/\sigma=2.5, 5.0$ and $7.5$,
as specified in Sec.~\ref{sec:ModelForceField} in SM.
As explained in detail in Ref.~\cite{NahaliRosa2018},
these values produce an efficient exploration across the relevant length and time scales of the polymer solutions (see Table~\ref{tab:RelevantLengthTimeScales} in SM),
from $\approx \xi$ (the so called {\it correlation length}~\cite{RubinsteinColbyBook}, marking the transition from solvent- to polymer-dominated physics) to $\approx 2 d_T$ (the so called {\it tube diameter}~\cite{DoiEdwardsBook}, marking the next transition to topology-dominated physics).

We study the static and kinetic properties of polymer chains and nanoprobes using fixed-volume molecular dynamics simulations with implicit solvent and periodic boundary conditions.
By defining $V$ the volume of the simulation box {\it accessible} to the polymers, the overall monomer density of the system $\rho \equiv (NM)/V$ is fixed to $0.3 / \sigma^3$.
This set-up is consistent with the one studied in Ref.~\cite{NahaliRosa2018} and we refer the reader to that publication and to SM for additional details.
MD simulations were performed using the LAMMPS package~\cite{LammpsPlimptonJCP1995} using a velocity Verlet algorithm, in which all beads and nanoprobes are weakly coupled to a Langevin heat bath with a local damping constant $\Gamma = 0.5 \tau_{\rm MD}^{-1}$
where
$\tau_{\rm MD} = \sigma(m / \epsilon)^{1/2}$ is the MD Lennard-Jones time unit,
$\epsilon$ is the energy unit
and
$m$ is the conventional mass unit for both monomers and nanoprobes.
The integration time step is $\tau_{\rm int} = 0.006 \, \tau_{\rm MD}$.
The systems were run long enough for the chains static properties to reach a steady state and to move more than their own size (see Fig.~\ref{fig:msd_vs_Rg} in SM).

Similarly to the protocol by Smrek and Kremer~\cite{SmrekKremerPRL2017}, $M/2=40$ chains of our systems are driven out-of-equilibrium by coupling their monomers to a ``hotter'' thermostat with temperature $T_h > T_c=\epsilon/k_B$ (see Sec.~\ref{sec:ModelForceField} in SM for the definition of energy scales) where $T_c$ is the temperature of the thermostat coupled to the remaining chains:
we name the polymers in the first group ``hot'' or ``active'' and the ones in the second ``cold'' or ``passive''.

By defining the ``reduced'' temperature gap $\Delta t \equiv \frac{T_h-T_c}{T_c} = \frac{T_h}{T_c}-1$,
we simulate systems with $T_h / T_c = 1.5, 2.0$ or $\Delta t = 0.5, 1.0$.
To gain physical insight, we compare the physical properties of these systems to the purely passive counterparts with homogeneous temperature $T_h = T_c = T = \epsilon/k_B$ or $\Delta t=0$.


%
\begin{figure}
\includegraphics[width=0.485\textwidth]{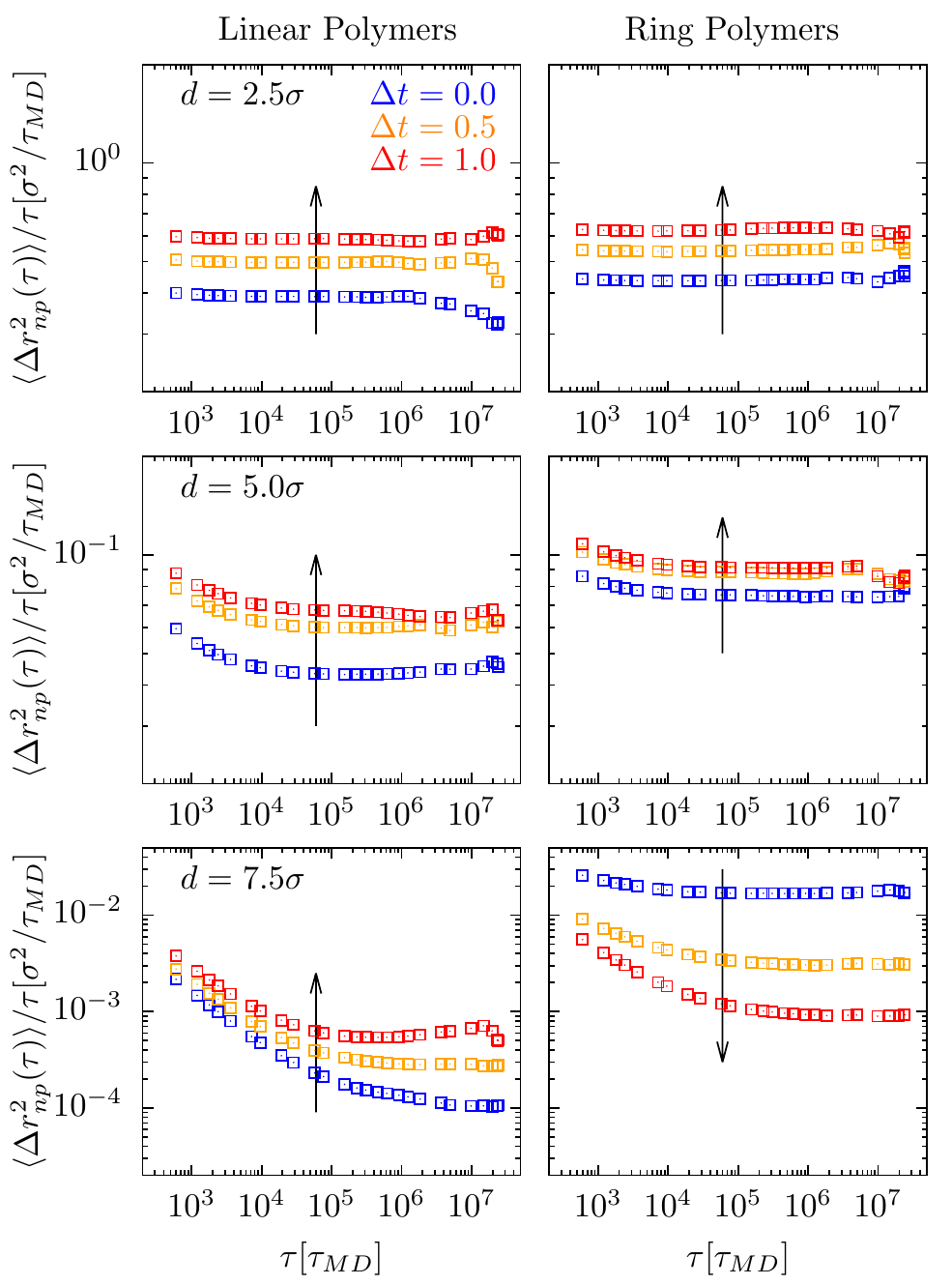} 
\caption{
Nanoprobe mean-square displacement (Eq.~\eqref{eq:DefineNpMSD}) per unit time, $\langle \Delta r_{\rm np}^2 (\tau) \rangle / \tau$. 
Results for increasing nanoprobe diameters $d$ (from top to bottom) and for linear chains (l.h.s. panels) 
{\it vs.} rings (r.h.s. panels). 
Different colors are for different reduced temperature gaps $\Delta t$ (see caption), whose increasing in magnitude is indicated by the corresponding arrow direction.
All systems have reached the proper diffusive regime as manifested by corresponding plateaus.
}
\label{fig:npMSD}
\end{figure}

{\it Results} --
We characterize the stochastic motion of the ensemble of $N_{\rm np}$ nanoprobes in each polymer solution
by introducing the mean-square displacement as a function of the lag-time $\tau$,
\begin{equation}\label{eq:DefineNpMSD}
\langle \Delta r_{\rm np}^2(\tau) \rangle \equiv \frac1{N_{\rm np}} \sum_{i=1}^{N_{\rm np}} \, \Delta r_{{\rm np}, i}^2(\tau)
\end{equation}
where
\begin{eqnarray}\label{eq:DefineNpMSD-TimeAv}
\Delta r_{{\rm np}, i}^2(\tau)
& \equiv & \left. \Delta r_{{\rm np}, i}^2({\mathcal T}, \tau) \, \right|_{{\mathcal T} \rightarrow \infty} \nonumber\\
& = & \left. \frac{1}{\mathcal T-\tau} \int_{0}^{\mathcal T-\tau}\left({\vec r}_i(t+\tau) - {\vec r}_i(t) \right)^2 dt \, \right|_{{\mathcal T} \rightarrow \infty}
\end{eqnarray}
is the time average mean-square displacement for the $i$-th nanoprobe
of spatial coordinates $\vec r_i(t)$ at time $t$
(see Sec.~\ref{sec:DefineObservables} in SM for additional details on these quantities).

The behaviors of $\langle \Delta r_{\rm np}^2(\tau) \rangle$ in solutions of linear chains and ring polymers, for different reduced temperatures $\Delta t$ and different nanoprobe diameters $d$ are summarized in Fig.~\ref{fig:npMSD}.
By comparing $d$ to the tube diameter $d_T \approx 4.3\sigma$ of the fully passive solutions (see Table~\ref{tab:RelevantLengthTimeScales} in SM), we may clearly identify two regimes:

(i) $d \lesssim d_T$, Fig.~\ref{fig:npMSD}, top and middle row.
Here the two thermostats produce similar effects regardless of the polymer architecture,
simply the nanoprobes display a higher effective temperature with respect to the fully passive case (Table~\ref{tab:NpEffectiveTs} in SM) and hence diffuse faster.
At the same time nanoprobe diffusion in ring solutions is always faster than in linear ones,
similarly to previous~\cite{NahaliRosa2016,NahaliRosa2018,GeRubinsteinMacromolecules2017} reports for passive systems.
As a marginal yet less intuitive effect, after subtracting the effect of the thermal speed-up (see Fig.~\ref{fig:npMSD-Ratios} in SM for ratios of nanoprobe mean-square displacements in ring {\it vs.} linear solutions at fixed $\Delta t$) we isolate a slow-down of the nanoprobes at increasing $\Delta t$:
without going into the details of it, we are tempted to ascribe this effect to the dependence of entanglements on chain flexibility (see our comment in the caption of Fig.~\ref{fig:npMSD-Ratios} in SM).

\begin{figure}
\includegraphics[width=1.0\columnwidth]{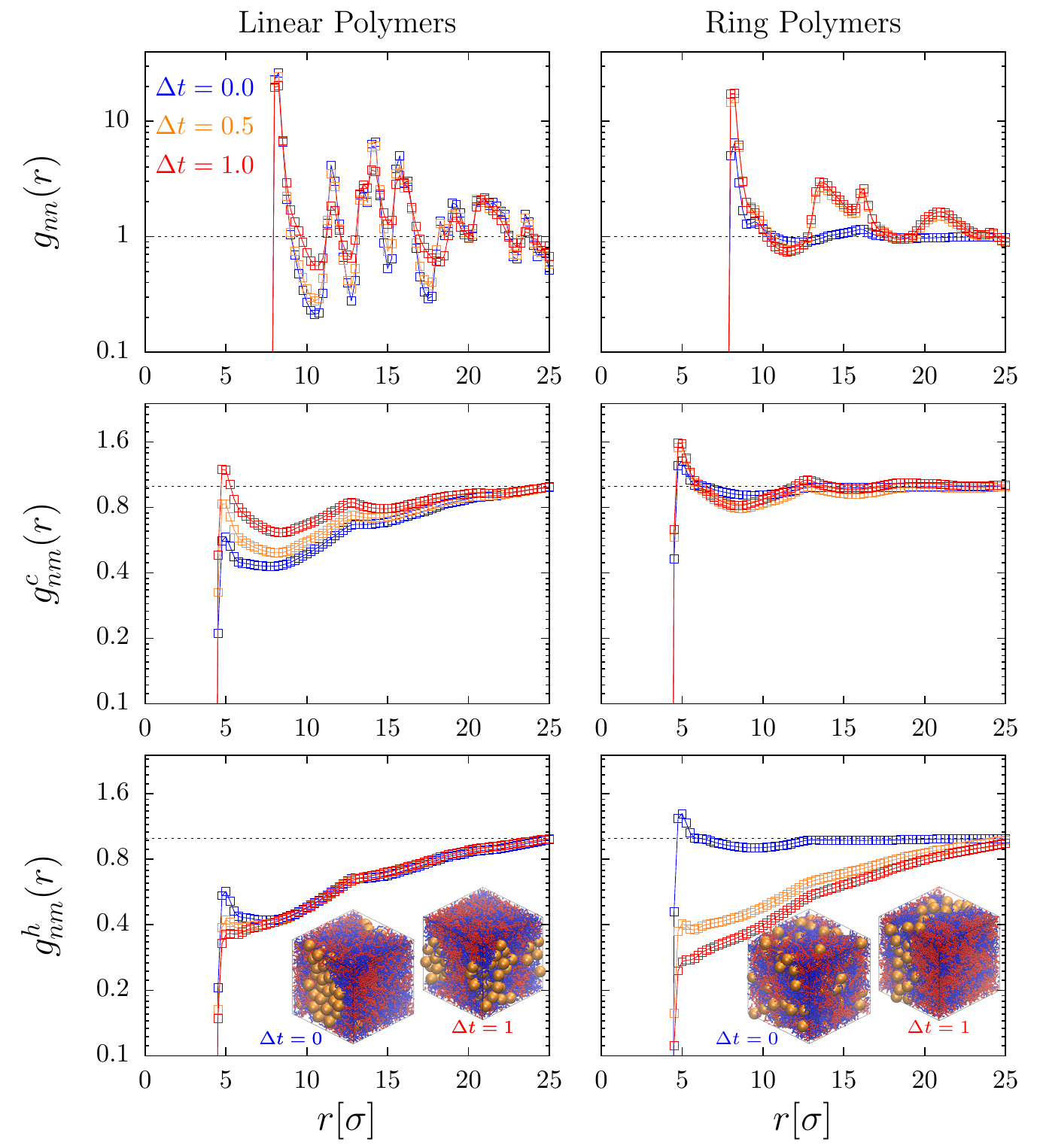}
\caption{
Nanoprobe-nanoprobe ($g_{\rm nn}(r)$) and nanoprobe-monomer ($g_{\rm nm}^{\rm c}(r)$ and $g_{\rm nm}^{\rm h}(r)$) pair correlation functions
for nanoprobes of diameter $d/\sigma=7.5$. 
The superscripts indicate that the functions have been evaluated by separating the contributions of monomers coupled to the cold (c) or the hot (h) thermostat.
Color code is as in Fig.~\ref{fig:npMSD}.
The insets in the bottom-row panels show typical configurations for $\Delta t=0.0$ and $\Delta t=1.0$ with cold/hot chains in blue/red and nanoprobes in yellow.
}
\label{fig:npGR_largest}
\end{figure}

(ii) $d \gtrsim d_T$, Fig.~\ref{fig:npMSD}, bottom row.
The situation for the largest nanoprobes is quite different:
while still agreeing with the reported~\cite{NahaliRosa2016,NahaliRosa2018,GeRubinsteinMacromolecules2017} observation that in passive ($\Delta t=0$) systems nanoprobes move faster in ring solutions than in solutions of linear chains (here $\approx 100$ times faster in the free diffusion regime), this discrepancy is significantly {\it reduced} upon driving the corresponding systems out of equilibrium (orange and red symbols).
The diffusion in ring solutions drops more than one order of magnitude while increasing of $\approx 5$ times in solutions of linear chains with increasing temperature gap.
We discuss these results in terms of the density fluctuations around the large nanoprobes and consider separately the three radial pair correlation functions, $g_{\rm nn}(r)$, $g_{\rm nm}^c(r)$ and $g_{\rm nm}^h(r)$, for nanoprobe {\it vs.} nanoprobe and nanoprobes {\it vs.} (cold/hot) monomers. 
In linear solutions (Fig.~\ref{fig:npGR_largest}, l.h.s. panels) large nanoprobes cluster at any temperature difference (including the equilibrium case $\Delta t =0$) in contrast to rings.
This entropic effect, consequent on the different chain architecture and whose details will be explored elsewhere,
naturally slows down the diffusion owing to steric effects.
However, as the $\Delta t$ increases, the effective temperature of the nanoprobes grows due to the heat transfer in the system, opposing clustering and letting nanoprobes to `fluidize'
(as manifested by the progressive levelling of the secondary peaks of $g_{\rm nn}(r)$ with increasing $\Delta t$ and the corresponding increasing of $g_{\rm nm}^c(r)$).
These effects are also visible in the two configurations for $\Delta t=0.0$ and $\Delta t=1.0$ shown in the bottom l.h.s. panel in Fig.~\ref{fig:npGR_largest}.
In contrast (Fig.~\ref{fig:npGR_largest}, r.h.s. panels) to the linear case, nanoprobes are well interspersed in passive ($\Delta t=0$) ring solutions while they exhibit clustering at $\Delta t >0$.
This clustering (evident in the two conformations for $\Delta t=0.0$ and $\Delta t=1.0$ in the bottom r.h.s. panel in Fig.~\ref{fig:npGR_largest}) is driven by {\it non-equilibrium} phase separation~\cite{GrosbergJoanny2015,weberweberfreyPRL2016,SmrekKremerPRL2017,SmrekEntropy2018} between the hot rings and the nanoprobes (see the corresponding depletion hole in $g_{\rm nm}^h(r)$)
and is confirmed (see Table~\ref{tab:NpEffectiveTs} in SM) by the nanoprobe lower effective temperature in rings in comparison to linear polymers.
As argued recently in~\cite{SmrekEntropy2018}, in non-equilibrium phase separation the unlike species minimize contact interface in order to decrease the total entropy production rate in the system.
Notice that, consistently with the other results for smaller nanoprobes (Fig.~\ref{fig:npMSD}), such non-equilibrium effects are considerably reduced (if not absent at all) when $d\lesssim d_T$ (see Fig.~\ref{fig:npGR} in SM for a detailed comparison).

\begin{figure}
\includegraphics[width=0.48\textwidth]{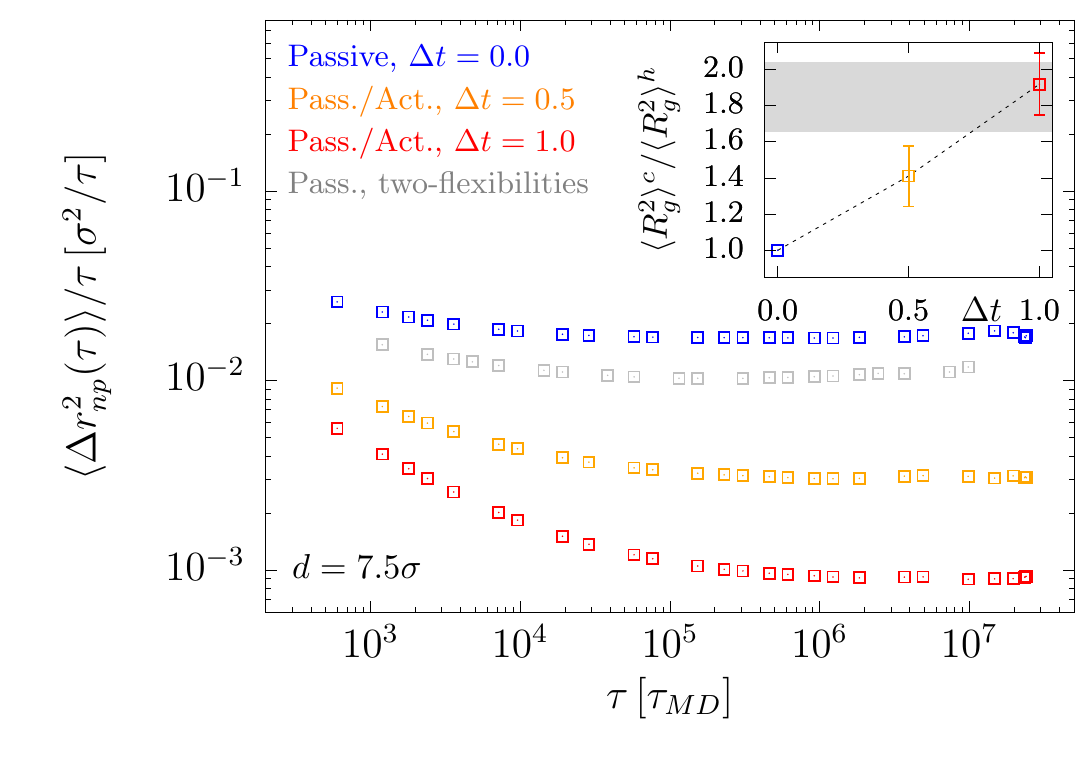}
\caption{
Mean-square displacement (Eq.~\eqref{eq:DefineNpMSD}) of large nanoprobes ($d=7.5\sigma$) in ring solutions and per unit time, $\langle \Delta r_{\rm np}^2 (\tau) \rangle / \tau$:
comparison of results for active-passive mixtures ($\Delta t > 0$, orange and red symbols), fully passive solutions with homogeneous polymer stiffness ($\Delta t=0$, blue symbols) and fully passive solutions with two polymer stiffnesses (gray symbols).
(Inset)
Corresponding ratios, $\langle R_g^2 \rangle^c / \langle R_g^2 \rangle^h$, of mean-square gyration radii (see Sec.~\ref{sec:DefineObservables} in SM for definitions) of ring polymers coupled to cold/hot thermostats, error bars are for standard deviations of the mean.
The grey strip corresponds to the ratio calculated for ``stiffer {\it vs.} less stiff'' polymers in the fully passive model with two chain flexibilities (see main text and Sec.~\ref{sec:MDruns} in SM for details).
}
\label{fig:avRg2-with-npMSD}
\end{figure}

These two points explain the contrasting trends in nanoprobe diffusion suspended in rings in comparison to linear chains.
However, as seen in Figs.~\ref{fig:R2vsN-Over-N} and~\ref{fig:R2vsN} in SM, the two populations of linear chains and rings react to non-equilibrium conditions quite differently:
compared to their passive counterparts, linear chains (both cold and hot) shrink while only hot rings do that and cold ones swell.
The shrinking arises from effectively higher temperature and hence flexibility.
In linear chains, where no permanent TC's exist this leads to shrinking of all chains as also cold ones have higher effective temperature than in equilibrium.
The contrasting behavior of rings results from the permanent TC's: unknotted rings oppose shrinking as that would lead to increase of knotted states prohibited by the TC's.
As shown below on equivalent equilibrium systems, the competition of the entropy loss from TC and the entropy gain from shrinking of more flexible chains yields the contrasting behavior of rings.
It is legitimate to suspect then that this ``asymmetry'' of the single-chain size in the two populations might trigger the nanoprobe dynamic behavior seen in Fig.~\ref{fig:npMSD}.

That this is not sufficient, {\it i.e.} that genuine non-equilibrium conditions are at the basis of the reported nanoprobe dynamics, can be demonstrated by the following argument.
We perform an additional simulation for polymer solutions with large ($d/\sigma=7.5$) nanoprobes and under purely passive conditions
and by fixing the stiffness of $50\%$ of the ring population to half of the original value (see Sec.~\ref{sec:MDruns} in SM for details).
Under these conditions, the average single-chain size is different for the two populations and matches the observed swelling of cold {\it vs.} hot rings in active-passive mixtures for $\Delta t=1.0$
(see inset in Fig.~\ref{fig:avRg2-with-npMSD}, showing the ratios of the steady-state polymer mean-square gyration radii, $\langle R_g^2 \rangle$, for the two polymer populations).
By comparing the nanoprobe mean-square displacement per unit time, $\langle\Delta r_{\rm np}^2(\tau)\rangle / \tau$, between this case and the former set-up's (see Fig.~\ref{fig:avRg2-with-npMSD}, main panel)
we see that the swelling observed in half of the chain population does not account for the nanoprobe slowdown seen in active-passive mixtures.

\begin{figure}
\includegraphics[width=\columnwidth]{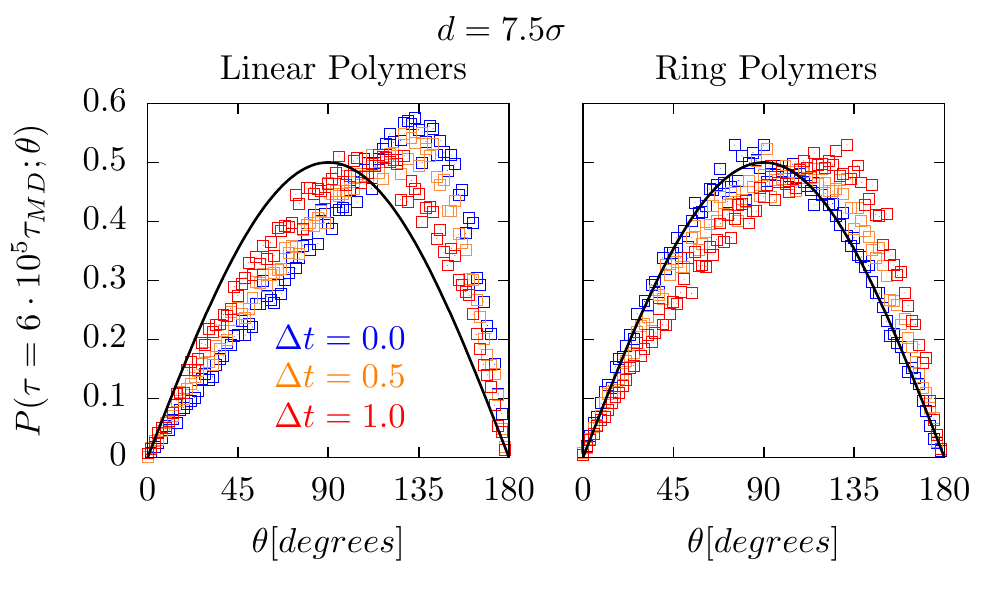} 
\caption{
\label{fig:pTheta}
Probability distribution functions, $P(\tau; \theta)$ (Eq.~\eqref{eq:AngleCorrelFunct}), of the angle $\theta$ between oriented spatial displacements of nanoprobes of diameter $d/\sigma=7.5$ and lag-time $\tau/\tau_{\rm MD}=6\cdot10^{5}$.
Color code is as in the rest of the paper, with different colors corresponding to reduced temperatures $\Delta t = 0.0, 0.5, 1.0$.
The black solid line is the function $P(\tau; \theta) = \frac12\sin\theta$ for randomly oriented vectors.
}
\end{figure}

Dynamic correlations in the motion of single nanoprobes can be characterized
by introducing the correlation function of the angle $\theta$ between oriented nanoprobe spatial displacements separated by lag-time
$\tau$~\cite{ValetRosa2014,NahaliRosa2018,PapaleRosaPhysBiol2019}:
\begin{align}
\label{eq:AngleCorrelFunct}
P(\tau; \theta) = \left \langle \theta - \cos^{-1}(\vec{u}(t+\tau)\cdot\vec{u}(t) )\right\rangle
\end{align}
where $\vec{u}(t) = (\vec r(t+\tau) - \vec r(t))/|\vec r(t+\tau) - \vec r(t)|$  is the (normalized) vector spatial displacement of the generic nanoprobe from time $t$ to $t+\tau$
and the brackets mean the same ensemble average defined for Eq.~\eqref{eq:DefineNpMSD}.
For randomly oriented displacements $P(\tau; \theta) = \frac12 \sin\theta$, any deviation from the null distribution being indicative of spatial correlations.
The distributions $P(\tau; \theta)$ at a large lag-time $\tau$ ({\it i.e.}, when the nanoprobes are already diffusive, see Fig.~\ref{fig:avRg2-with-npMSD}) are shown in Fig.~\ref{fig:pTheta} for large nanoprobes.
In both linear and ring solutions, nanoprobes exhibit correlations favoring backward displacements~\cite{NahaliRosa2016}, but with contrasting dependence on the temperature gap. While in linear solutions the nanoprobes become more and more unbiased at increasing $\Delta t$, the nanoprobes in ring solutions are unbiased in passive melts~\cite{NahaliRosa2016} while displaying directional correlations at high $\Delta t$, consistent with the proposed explanation of the diffusive data shown in Fig.~\ref{fig:npMSD}.
Instead, smaller nanoprobes display always unbiased distributions at large lag-times (see Fig.~\ref{fig:pTheta_all} in SM showing examples of distributions for all nanoprobe diameters and shorter lag-times).
Consistent with a recent study~\cite{NahaliRosa2018} on nanoprobe dynamics in entangled passive polymer melts, the presence of such correlations implies
(i)
that the motion of nanoprobes is spatially heterogeneous with (see Eqs.~\eqref{eq:DefineNpMSD} and~\eqref{eq:DefineNpMSD-TimeAv} for definitions) $\lim_{\mathcal T\rightarrow\infty} \Delta r_{\rm np}^2 ({\mathcal T}; \tau)$ not converging, in general, to $\langle \Delta r_{\rm np}^2 (\tau) \rangle$ (see Fig.~\ref{fig:SH} in SM)
and
(ii)
that the (so called~\cite{BerthierKobPRL2007,MetzlerVattulainen2016,MichielettoNahaliRosa2017} van-Hove, see Sec.~\ref{sec:NanoprobeDynamics} in SM) distribution functions, $P(\tau; \Delta x)$, of the Cartesian components of generic nanoprobes displacements $\vec r(\tau+t) - \vec r(t)$ from time $t$ to $t + \Delta t$
exhibit characteristic non-Gaussian heavy tails (see Fig.~\ref{fig:DisplsPDF} in SM).

{\it Discussion and conclusions} --
Active matter~\cite{BechingerRMP2016} and, in particular, active polymers~\cite{WinklerGompper2017} are an emergent research field in modern soft matter.
By employing the two-diffusivities dynamic model introduced and studied in the recent works~\cite{weberweberfreyPRL2016,SmrekKremerPRL2017,GrosbergJoanny2015,Ilker2021longtime},
we have shown that the way nanoprobes diffuse in active-passive polymer mixtures depend on the architecture of the chains.
In linear solutions activity disrupts the clustering of nanoprobes seen under purely passive conditions,
while in ring solutions the tendency is quite the opposite with nanoprobes separating away from the active polymer component (Fig.~\ref{fig:npGR_largest}).
Overall, this leads to the ``counterintuitive'' effect that activity accelerates nanoprobes in linear solutions but decelerates them in ring ones.

Following works~\cite{SmrekKremerPRL2017,SmrekEntropy2018}, we calculate the ``temperature asymmetry'' order parameters (see Table~\ref{tab:NpEffectiveTs} in SM) of the hot chains with respect to the nanoprobes ($\chi_{\rm np}^h$) and the cold chains ($\chi_c^h$).
We find that $\chi_{\rm np}^h({\rm ring})>\chi_{\rm np}^c({\rm linear})$ which is indeed consistent~\cite{SmrekKremerPRL2017} with nanoprobe clustering and separating from the active ring polymer component.
At the same time, in spite of the fact that $\chi_{\rm np}^h < \chi_c^h$, we report that we do not find evidence for {\it phase separation} of polymer chains as reported in~\cite{SmrekKremerPRL2017,SmrekEntropy2018,IlkerJoannyPRR2020}:
this may be due, {\it in primis}, to the fact that the polymer systems used here are more dilute than the ones employed in those previous work.
Nonetheless, we see that activity has still some non trivial effect on the chains based, once again, on architecture:
both linear chain populations reduce their size as a consequence of the activity, while hot rings crumple and cold ones swell (Fig.~\ref{fig:R2vsN-Over-N} in SM).

We conclude with a possible connection to the biophysics of interphase chromosomes.
It has been hypothesized~\cite{RosaPLOS2008,RepProgPhys_RingsChromosomes_2014} that the microscopic topological state of chromatin (the linear fiber made of DNA and proteins which constitute the primary component of eukaryotic chromosomes~\cite{AlbertsBook}) in the cell nucleus is akin to a melt of rings.
Differently from standard polymer melts and because of undergoing energy-consuming processes like, for instance, loop extrusion~\cite{GoloborodkoBJ2016,Stam2019} or transcription~\cite{Zidovska2020}, a certain fraction of the chromatin inside the cell is constantly maintained out of equilibrium~\cite{ganaiNAR2014}.
Our results demonstrate that the motion of nanoprobes of diameter of the order of the chromatin mesh size ($\approx 300$nm~\cite{RosaPLOS2008,ValetRosa2014}) or larger is deeply influenced by the thermal state of the chromatin fiber:
in a typical~\cite{HameedShivashankarPlosOne2012} microrheology experiment performed in the nucleus,
nanoprobes are expected to separate from the more active chromatin regions by forming clusters within the inactive ones.
Last but not least our results suggest that chromatin activity, and not only chromatin conformation as usually~\cite{BroekPNAS2008,AmitaiBJ2018,LizanaPRRes2021} pointed out, is arguably controlling the dynamics of DNA-regulatory proteins towards their target sequences in the cell nucleus.

{\it Acknowledgements} --
JS and AR acknowledge networking support by the COST Action CA17139 (EUTOPIA). JS acknowledges the support from the Austrian Science Fund (FWF) through the Lise-Meitner Fellowship \mbox{No.~M 2470-N28}. JS is grateful for the computational time at Vienna Scientific Cluster. AR and AP acknowledge computational resources from SISSA HPC-facilities.

\bibliography{biblio}

\clearpage

\widetext
\clearpage
\begin{center}
\textbf{\Large Nanorheology of active-passive polymer mixtures is topology-sensitive \\ \vspace*{1.5mm} -- Supplemental Material --} \\
\vspace*{5mm}
Andrea Papale, Jan Smrek, Angelo Rosa
\vspace*{10mm}
\end{center}
\balancecolsandclearpage

\setcounter{equation}{0}
\setcounter{figure}{0}
\setcounter{table}{0}
\setcounter{page}{1}
\setcounter{section}{0}
\makeatletter
\renewcommand{\theequation}{S\arabic{equation}}
\renewcommand{\thefigure}{S\arabic{figure}}
\renewcommand{\thetable}{S\arabic{table}}
\renewcommand{\thesection}{S\arabic{section}}

\makeatletter
\@fpsep\textheight
\makeatother

\section{Model and methods: details}\label{sec:ModelMethods-MoreDetails}
In this Section, we give additional details related to
the polymer/nanoprobe model used in this work (Sec.~\ref{sec:ModelForceField})
and
the computational effort required for the simulations (Sec.~\ref{sec:MDruns}).
Then, we conclude (Sec.~\ref{sec:DefineObservables}) by describing the mathematical details beyond the calculation of some observables considered in this work.

\subsection{Computational model for polymers and nanoprobes}\label{sec:ModelForceField}
Polymer-polymer interactions consist of the following three terms:
\begin{itemize}
\item[(i)]
The potential energy term accounting for monomer-monomer excluded volume interactions,
which is expressed by 
the shifted and truncated Lennard-Jones (LJ) function:
\begin{equation}\label{eq:ULJ}
U_{\rm LJ}(r) = \begin{cases} 4\epsilon\left[\left(\frac{\sigma}{r}\right)^{12}-\left(\frac{\sigma}{r}\right)^{6}+\frac{1}{4}\right]  & r\leq r_c \\
0 & r>r_c \end{cases} \, .
\end{equation}
Here, $r$ is the spatial distance between monomers and the chosen cut-off distance $r_c / \sigma= 2^{1/6}$ ensures that only purely repulsive monomer-monomer interactions are effectively taken into account. 
The parameters
$\epsilon$ 
and
$\sigma$ fix the energy and length scales units, respectively.
\item[(ii)]
The bond potential between monomers which are nearest-neighbours along the same polymer chain,
which is expressed by the so called finitely extensible non-linear elastic potential (FENE):
\begin{equation}\label{eq:Ufene}
U_{\rm FENE}(r) = \begin{cases} -\frac12 \, \kappa_{\rm FENE} \, R^2_0 \, \log \left(1-\left(r/R_0\right)^2\right) & r \leq R_0 \\
\infty & r > R_0 \end{cases} \, .
\end{equation}
Here, $\kappa_{\rm FENE} \, \sigma^2 / \epsilon = 30$ is the spring constant and $R_0 / \sigma =1.5$ is the maximum extension of the elastic FENE bond.
\item[(iii)]
The bending energy term controlling polymer stiffness,
which is expressed by the following function: 
\begin{equation}\label{eq:Ubend}
U_{\rm bend}(\theta) = \kappa_{\rm bend} \left ( 1 - \frac{ (\vec r_{i+1} - \vec r_i) \cdot (\vec r_i-\vec r_{i-1}) }{ |\vec r_{i+1} - \vec r_i| \, |\vec r_i-\vec r_{i-1}| } \right) \, . 
\end{equation}
Here, $\vec r_i$ is the coordinate of the $i$-th monomer along each given chain, numbered from one of the termini (for linear chains) or from an arbitrarily chosen monomer (for rings).
In the latter case, periodic boundary conditions along the ring are tacitly assumed. 
The bending constant $\kappa_{\rm bend} /\epsilon = 5$,  
corresponding to a Kuhn~\cite{DoiEdwardsBook,RubinsteinColbyBook} segment $\ell_K / \sigma = 10$~\cite{RosaPLOS2008}. 
\end{itemize}

The polymer solutions are accompanied by the presence of nanoprobes of different diameters. 
In order to model the nanoprobe-nanoprobe and nanoprobe-polymer interactions, 
we have resorted to the phenomenological expressions introduced by Everaers and Ejtehadi~\cite{EveraersEjtehadi2003} and employed in previous works~\cite{ValetRosa2014,NahaliRosa2016,NahaliRosa2018,PapaleRosaPhysBiol2019}.
In particular:
\begin{itemize}
\item[(iv)]
Nanoprobe-nanoprobe (nn) interactions are described by the expression:
\begin{equation}\label{eq:Unano-nano}
\begin{cases} U_{\rm nn}(r) = U^A_{\rm nn}(r)+U^R_{\rm nn}(r) & r\leq r_{\rm nn} \\
0 & r>r_{\rm nn} \end{cases} \, .
\end{equation}
$U_{\rm nn}^A(r)$ is the attractive contribution given by
\begin{equation}\label{eq:Unano-nano-attr}
U_{\rm nn}^A(r)=-\frac{A_{\rm nn}}{6}\left[\frac{2a^2}{r^2-4a^2}+\frac{2a^2}{r^2}+\text{ln}\left(\frac{r^2-4a^2}{r^2}\right)\right] \, ,
\end{equation}
while $U_{\rm nn}^B(r)$ is the repulsive term
\begin{eqnarray}\label{eq:Unano-nano-rep}
U_{\rm nn}^B(r)
& = & \frac{A_{\rm nn}}{37800}\frac{\sigma^6}{r}\left[\frac{r^2-14ar+54a^2}{\left(r-2a\right)^7} + \right. \nonumber\\
& & \left. \frac{r^2+14ar+54a^2}{\left(r+2a\right)^7} - 2\frac{r^2-30a^2}{r^7}\right] \, .
\end{eqnarray}
Here,
$A_{\rm nn} / \epsilon = 39.478$ and 
we consider non-sticky, athermal probe particles with diameters $d / \sigma \equiv 2a / \sigma =  2.5, 5.0, 7.5$
corresponding to fix $r_{\rm nn} / \sigma=3.08, 5.60, 8.08$.
As explained in great detail in Ref.~\cite{NahaliRosa2018} 
these nanoprobe diameters have been chosen because 
(a)
they are larger than the {\it correlation length}~\cite{RubinsteinColbyBook} $\xi / \sigma \approx 1.4$ of the polymer solution
while, at the same time,
(b)
they are able to span the entire range from below to above the estimated value $d_T / \sigma \approx 4.3$ of the {\it tube diameter}
(see Table~\ref{tab:RelevantLengthTimeScales} for an overview of the physical property of the polymer solutions employed here).
In this way,
(a)
polymer effects on nanoprobe displacement dominate~\cite{GeRubinsteinMacromolecules2017} over thermal effects caused by the solvent
and
(b)
the role of entanglements on nanoprobe motion can be explored more systematically.
\item[(v)]
Finally, the monomer-nanoprobe (mn) interaction is accounted for by:
\begin{equation}\label{eq:Unano-polymer}
\begin{cases}
U_{\rm mn}(r) =\frac{2a^3\sigma^3A_{\rm mn}}{9\left(a^2-r^2\right)^3}
\left[1-\frac{5a^6+45a^4 r^2+63a^2 r^4+15r^6}{15\left(a-r\right)^6\left(a+r\right)^6}\right] & r\leq r_{\rm mn} \\
0 & r>r_{\rm mn} \end{cases} \, 
\end{equation}
where $A_{\rm mn} / \epsilon = 75.358$ 
and $r_{\rm mn}/\sigma = 2.11, 3.36, 4.61$.
\end{itemize}
%

\begin{table}
\begin{tabular}{cc}
\hline
\hline
\\
Quantity & Value \\
\hline
\\
Correlation length, $\xi / \sigma$ & 1.4 \\
Entanglement length, $L_e / \sigma$ & 11.0 \\
Tube diameter, $d_T / \sigma$ & 4.3 \\
Entanglement time, $\tau_e / \tau_{\rm MD}$ & 490.0 \\
\hline
\hline
\end{tabular}
\caption{
\label{tab:RelevantLengthTimeScales}
List of relevant length and time scales describing the microscopic properties of the polymer solution:
(i)
The correlation length, $\xi$, is defined as the average spatial distance from a monomer on one chain to the nearest monomer on another chain~\cite{RubinsteinColbyBook}
and it is a measure of the packing of the solution;
(ii)
The entanglement length, $L_e$, can be defined as the contour length along a single chain which spans between close-by entanglement points in the solution~\cite{EveraersScience2004,UchidaEveraersJCP2008};
(iii)
The tube diameter, $d_T \approx \sqrt{\ell_K L_e}$, measures the average span in length between close entanglement points along the same chain~\cite{EveraersScience2004,UchidaEveraersJCP2008};
(iv)
The entanglement time, $\tau_e$, is the average time for a monomer to explore by random motion a portion of the solution of linear size $=d_T$~\cite{EveraersScience2004,UchidaEveraersJCP2008}.
}
\end{table}

\subsection{Molecular dynamics runs}\label{sec:MDruns}
As explained in the main text, we have performed Langevin molecular dynamics for a polymer system made of $M=80$ chains of $N=500$ beads each and $N_{\rm np} = 100$ nanoprobes dispersed in the solution.
Simulations were performed by using the LAMMPS package~\cite{LammpsPlimptonJCP1995}.
Half of the chains are coupled to a thermostat with ``room'' temperature $T_c = T\equiv\epsilon/k_B$ ($k_B$ being the Boltzmann constant) and the other half are coupled to a ``hotter'' thermostat with temperature $T_h / T_c > 1$.
The nanoprobes are always coupled to the cold thermostat.
By defining the ``reduced'' temperature gap $\Delta t \equiv T_h / T_c-1$,
we have considered systems with $T_h / T_c=1.5$ or $\Delta t =0.5$ and $T_h / T_h = 2.0$ or $\Delta t =1.0$.
Then, we have compared the properties of these systems to those for ``purely passive'' solutions with $T_h / T_c = 1.0$ or $\Delta t =0.0$.

Polymers/nanoprobes mixtures are prepared and then let equilibrate under purely passive conditions according to the protocol described in detail in Ref.~\cite{NahaliRosa2018}.
Starting from these equilibrated systems, half of the chains are then driven out of equilibrium by the coupling to the hot thermostat.
The total length of each MD run is $\approx4\cdot 10^9$ integration time steps $\tau_{\rm int}$ (with our choice $\tau_{\rm int} / \tau_{\rm MD} = 0.006$, this is equivalent to about $\approx2.4\cdot10^7$ MD Lennard-Jones time units).
System configurations are sampled each $10^5 \, \tau_{\rm int} = 600 \, \tau_{\rm MD}$:
in order to remove possible artifacts due to the initial preparation of the samples, 
all the analyses reported in this work have been performed after discarding the first $5\cdot10^7 \, \tau_{\rm int} = 3\cdot 10^5 \, \tau_{\rm MD}$ of each trajectory.
For completeness and in order to investigate smaller time scales, we have also performed additional runs of total length $\approx 2 \cdot 10^6 \, \tau_{\rm int} = 1.2\cdot 10^4\,\tau_{\rm MD}$ with reduced sampling time of $100 \, \tau_{\rm int} = 0.6\,\tau_{\rm MD}$.

\begin{table*}
\begin{tabular}{cccccccccccccc}
\hline
\hline\scriptsize
\\
& & & \multicolumn{5}{c}{Linear Polymers} & & \multicolumn{5}{c}{Ring Polymers} \\
\cline{4-8} \cline{10-14} \\
$d/\sigma$ & $\Delta t$ & & $\langle T_{\rm np} \rangle$ & $\langle T_{\rm ch} \rangle^c$ & $\langle T_{\rm ch} \rangle^h$ & $\chi_{\rm np}^h$ & $\chi_c^h$ & & $\langle T_{\rm np} \rangle$ & $\langle T_{\rm ch} \rangle^c$ & $\langle T_{\rm ch} \rangle^h$ & $\chi_{\rm np}^h$ & $\chi_c^h$ \\
\hline
\\
& \scriptsize{$0.0$} & & \scriptsize{$1.005 \pm 0.073$} & \scriptsize{$1.000 \pm 0.005$} & \scriptsize{$1.000 \pm 0.005$} & \scriptsize{$\approx 5\cdot10^{-2}$} & \scriptsize{$\lesssim10^{-3}$} & & \scriptsize{$1.030 \pm 0.082$} & \scriptsize{$1.000 \pm 0.005$} & \scriptsize{$1.000 \pm 0.005$} & \scriptsize{$\approx3\cdot10^{-2}$} & \scriptsize{$\lesssim10^{-3}$} \\
\scriptsize{$2.5$} & \scriptsize{$0.5$} & & \scriptsize{$1.200 \pm 0.100$} & \scriptsize{$1.067 \pm 0.006$} & \scriptsize{$1.432 \pm 0.007$} & \scriptsize{$0.19 \pm 0.02$} & \scriptsize{$0.342 \pm 0.004$} & & \scriptsize{$1.199 \pm 0.094$} & \scriptsize{$1.059 \pm 0.006$} & \scriptsize{$1.438 \pm 0.008$} & \scriptsize{$0.20 \pm 0.02$} & \scriptsize{$0.358 \pm 0.004$} \\
& \scriptsize{$1.0$} & & \scriptsize{$1.376 \pm 0.119$} & \scriptsize{$1.127 \pm 0.007$} & \scriptsize{$1.870 \pm 0.011$} & \scriptsize{$0.36 \pm 0.03$} & \scriptsize{$0.658 \pm 0.008$} & & \scriptsize{$1.359 \pm 0.112$} & \scriptsize{$1.122 \pm 0.007$} & \scriptsize{$1.877 \pm 0.010$} & \scriptsize{$0.38 \pm 0.03$} & \scriptsize{$0.672 \pm 0.008$} \\ 
\\
& \scriptsize{$0.0$} & & \scriptsize{$1.009 \pm 0.087$} & \scriptsize{$1.000 \pm 0.006$} & \scriptsize{$1.000 \pm 0.006$} & \scriptsize{$\approx 1\cdot10^{-2}$} & \scriptsize{$\lesssim10^{-3}$} & & \scriptsize{$1.020 \pm 0.083$} & \scriptsize{$1.000 \pm 0.006$} & \scriptsize{$1.000 \pm 0.006$} & \scriptsize{$\approx2\cdot10^{-2}$} & \scriptsize{$\lesssim10^{-3}$} \\
\scriptsize{$5.0$} & \scriptsize{$0.5$} & & \scriptsize{$1.191 \pm 0.105$} & \scriptsize{$1.069 \pm 0.006$} & \scriptsize{$1.431 \pm 0.008$} & \scriptsize{$0.20 \pm 0.02$} & \scriptsize{$0.339 \pm 0.004$} & & \scriptsize{$1.183 \pm 0.104$} & \scriptsize{$1.060 \pm 0.006$} & \scriptsize{$1.441 \pm 0.009$} & \scriptsize{$0.22 \pm 0.02$} & \scriptsize{$0.360 \pm 0.004$} \\ 
& \scriptsize{$1.0$} & & \scriptsize{$1.348 \pm 0.100$} & \scriptsize{$1.138 \pm 0.007$} & \scriptsize{$1.860 \pm 0.012$} & \scriptsize{$0.38 \pm 0.03$} & \scriptsize{$0.635 \pm 0.008$} & & \scriptsize{$1.323 \pm 0.087$} & \scriptsize{$1.124 \pm 0.007$} & \scriptsize{$1.877 \pm 0.011$} & \scriptsize{$0.42 \pm 0.03$} & \scriptsize{$0.669 \pm 0.008$} \\
\\
& \scriptsize{$0.0$} & & \scriptsize{$1.004 \pm 0.076$} & \scriptsize{$0.999 \pm 0.005$} & \scriptsize{$0.999 \pm 0.005$} & \scriptsize{$\approx4\cdot10^{-3}$} & \scriptsize{$\lesssim10^{-3}$} & & \scriptsize{$1.000 \pm 0.077$} & \scriptsize{$1.000 \pm 0.005$} & \scriptsize{$1.000 \pm 0.005$} & \scriptsize{$\lesssim10^{-3}$} & \scriptsize{$\lesssim10^{-3}$} \\
\scriptsize{$7.5$} & \scriptsize{$0.5$} & & \scriptsize{$1.172 \pm 0.093$} & \scriptsize{$1.075 \pm 0.006$} & \scriptsize{$1.424 \pm 0.008$} & \scriptsize{$0.21 \pm 0.02$} & \scriptsize{$0.324 \pm 0.004$} & & \scriptsize{$1.120 \pm 0.090$} & \scriptsize{$1.064 \pm 0.007$} & \scriptsize{$1.436 \pm 0.008$} & \scriptsize{$0.28 \pm 0.02$} & \scriptsize{$0.349 \pm 0.004$} \\
& \scriptsize{$1.0$} & & \scriptsize{$1.292 \pm 0.107$} & \scriptsize{$1.144 \pm 0.007$} & \scriptsize{$1.856 \pm 0.011$} & \scriptsize{$0.44 \pm 0.04$} & \scriptsize{$0.623 \pm 0.007$} & & \scriptsize{$1.163 \pm 0.095$} & \scriptsize{$1.125 \pm 0.007$} & \scriptsize{$1.875 \pm 0.011$} & \scriptsize{$0.61 \pm 0.05$} & \scriptsize{$0.667 \pm 0.008$} \\
\hline
\hline
\end{tabular}
\caption{
\label{tab:NpEffectiveTs}
Summary of average temperatures for nanoprobes ($\langle T_{\rm np} \rangle$) and for individual monomers of cold and hot chains ($\langle T_{\rm ch} \rangle^{c, h}$),
and corresponding ``temperature asymmetry'' order parameters
for hot chains with respect to
nanoprobes $(\chi_{\rm np}^h \equiv \frac{\langle T_{\rm ch} \rangle^h}{\langle T_{\rm np} \rangle}-1)$
and
for hot chains w.r.t. cold chains $(\chi_c^h \equiv \frac{\langle T_{\rm ch} \rangle^h}{\langle T_{\rm ch} \rangle^c}-1)$.
Temperatures are measured in the course of the simulations by the LAMMPS~\cite{LammpsPlimptonJCP1995} numerical engine used for this work (see Sec.~\ref{sec:MDruns}).
$d$ is the nanoprobe diameter and $\Delta t$ is the reduced temperature gap introduced in the system (see the main text and Sec.~\ref{sec:MDruns} for details).
}
\end{table*}

As shown in Fig.~\ref{fig:msd_vs_Rg}, the runs are long enough for the mean-square displacement to be above the squared gyration radius.
This is typically long enough to achieve the complete relaxation of polymer systems, see Ref.~\cite{HalversonKremerJCP2011-Dynamics}.
Table~\ref{tab:NpEffectiveTs} summarizes
the average temperature of the nanoprobes, $\langle T_{\rm np} \rangle$,
and the average temperatures of the monomers of cold and hot chains, $\langle T_{\rm ch}\rangle^{c,h}$, after the complete relaxation of the corresponding systems.
It reports also the corresponding values for the ``temperature asymmetry'' order parameters (see Ref.~\cite{SmrekKremerPRL2017})
for hot chains with respect to
nanoprobes $(\chi_{\rm np}^h \equiv \frac{\langle T_{\rm ch} \rangle^h}{\langle T_{\rm np} \rangle}-1)$
and
for hot chains w.r.t. cold chains $(\chi_c^h \equiv \frac{\langle T_{\rm ch} \rangle^h}{\langle T_{\rm ch} \rangle^c}-1)$.

In addition, we have performed a different run (of total length $=1.2\cdot10^7\,\tau_{\rm MD}$) for a fully passive systems of ring polymers and large nanoprobe with diameter $d/\sigma=7.5$.
The system and numerical details are as before:
the only exception is that now the bending stiffness of 50\% of the chain population is as before ($\kappa_{\rm bend}/\epsilon=5.0$, see Sec.~\ref{sec:ModelForceField} here) while the remaining 50\% of rings are twice more flexible with $\kappa_{\rm bend}/\epsilon = 2.5$.
By this protocol, the average chain sizes of the two populations of rings ``fit'' the sizes found for passive/active mixtures at $\Delta t=1.0$ (see inset in Fig.~\ref{fig:avRg2-with-npMSD} in the main paper).

\subsection{Observables and measured properties: definitions}\label{sec:DefineObservables}

\subsubsection{Single-chain structure}\label{sec:SingleChainStructure}
Let us define ${\mathcal O}_m(t)$, the value of the generic observable $\mathcal O$ referring to the $m$-th chain in the solution and evaluated at time step $t$ of a given MD run.
Its mean value, $\langle{\mathcal O}\rangle^{\rm c,h}$, is defined by the formula:
\begin{equation}\label{eq:MeanValueObs}
\langle {\mathcal O} \rangle^{\rm c, h} \equiv \frac1{M/2} \, \sum_{m=1}^{M/2}{\vphantom{\sum}}^{\rm c, h} \, \frac1{t_\ast} \int_{T-t_\ast}^T {\mathcal O_m(t)} \, dt \, ,
\end{equation}
where:
(a)
$t_\ast$ corresponds to the time scale above which chains, having diffused more than their own size, have reached the steady state (see Fig.~\ref{fig:msd_vs_Rg});
(b)
the subscripts on the brackets $\langle \cdot \rangle^{\rm c, h}$ mean that separate averages have been taken for the two chain populations coupled to the two thermostats.
In analogous manner, distinct averages have been considered in the case of chains with different flexibilities (Sec.~\ref{sec:MDruns}).

\begin{table}
\begin{tabular}{cccccccc}
\hline
\hline
\\
& & & \multicolumn{2}{c}{Linear Polymers} & & \multicolumn{2}{c}{Ring Polymers} \\
\cline{4-5} \cline{7-8} \\
$d/\sigma$ & $\Delta t$ & & $\langle R_g^2 \rangle^h$ & $\langle R_g^2 \rangle^c$ & & $\langle R_g^2 \rangle^h$ & $\langle R_g^2 \rangle^c$ \\
\hline
\\
& $0.0$ & & \multicolumn{2}{c}{$700.5 \pm 62.7$} & & \multicolumn{2}{c}{$167.7 \pm 9.2$} \\
$2.5$ & $0.5$ & & $488.2 \pm 32.5$ & $615.2 \pm 44.9$ & & $139.2 \pm 7.5$ & $170.4 \pm 10.5$ \\
& $1.0$ & & $391.7 \pm 31.4$ & $574.8 \pm 47.8$ & & $128.7 \pm 6.3$ & $181.5 \pm 10.0$ \\
\\
& $0.0$ & & \multicolumn{2}{c}{$690.9 \pm 52.1$} & & \multicolumn{2}{c}{$169.1 \pm 9.4$} \\
$5.0$ & $0.5$ & & $486.8 \pm 36.1$ & $583.5 \pm 41.4$ & & $138.1 \pm 7.6$ & $177.4 \pm 10.0$ \\
& $1.0$ & & $386.9 \pm 28.5$ & $576.6 \pm 43.0$ & & $132.2 \pm 7.6$ & $189.0 \pm 11.5$ \\
\\
& $0.0$ & & \multicolumn{2}{c}{$653.0 \pm 39.7$} & & \multicolumn{2}{c}{$174.7 \pm 10.4$} \\
$7.5$ & $0.5$ & & $451.6 \pm 33.2$ & $577.7 \pm 44.0$ & & $137.3 \pm 7.5$ & $193.6 \pm 12.6$ \\
& $1.0$ & & $368.6 \pm 30.3$ & $581.8 \pm 43.8$ & & $129.6 \pm 6.6$ & $248.5 \pm 9.4$ \\
\hline
\hline
\end{tabular}
\caption{
\label{tab:avRg2}
Mean-square gyration radii corresponding to the different chain population considered in this work.
The superscript ``$c$'' (respectively, ``$h$'') is for ``cold'' (resp. ``hot'') chains in the melt.
$d$ is the nanoprobe diameter and $\Delta t$ is the reduced temperature gap introduced in the system (see the main text and Sec.~\ref{sec:MDruns} for details).
$\Delta t=0$ is for classical passive melts and one single value is reported.
}
\end{table}

In this work, we have considered the following single-chain observables for which we have computed corresponding mean values according to the definition~\eqref{eq:MeanValueObs}:\\
(i)
The gyration radius of a polymer chain made of $N$ monomers, defined by:
\begin{equation}\label{eq:MeanSqGyrRad}
R_g^2(t) \equiv \frac1N \, \sum_{i=1}^N (\vec r_i(t) - \vec r_{\rm cm}(t))^2 \, ,
\end{equation}
where:
(a)
$\vec r_i(t)$ is the spatial position of the $i$-th monomer of the chain at time $t$;
(b)
$\vec r_{\rm cm}(t) \equiv \frac1N\sum_{i=1}^N\vec r_i$ is the position of the center of mass of the chain.
The mean-square gyration radii for the different chain populations are reproduced in Table~\ref{tab:avRg2}. \\
(ii)
The average square end-to-end distance between two monomers at given contour length separation $\ell\in [\sigma, (N-1)\sigma]$ along the chain, defined by:
\begin{equation}\label{eq:MeanSqEndToEnd}
R^2(\ell \equiv n\sigma; t) \equiv \frac1{N-n} \, \sum_{i=1}^{N-n} (\vec r_{i+n}(t) - \vec r_i(t))^2 \, ,
\end{equation}
where $\sigma$ is the average bond length (see Sec.~\ref{sec:ModelForceField}).
Definition~\eqref{eq:MeanSqEndToEnd} works for linear chains, the generalization to rings (where $\ell \in [\sigma, N\sigma/2]$) is obtained by taking into account the obvious periodicity along the contour length of the chain.

\subsubsection{Nanoprobe dynamics}\label{sec:NanoprobeDynamics}
To quantify the dynamics of single nanoprobes immersed in polymer solutions,
we introduce the mean-square displacement, $\Delta r_{{\rm np}, i}^2(\mathcal T; \tau)$, for the $i$-th nanoprobe ($i = 1, ..., N_{\rm np}=100$) as a function of the lag-time $\tau$ and the measurement time $\mathcal T$~\cite{MichielettoNahaliRosa2017,NahaliRosa2018,PapaleRosaPhysBiol2019}:
\begin{equation}\label{eq:npMSD}
\Delta r_{{\rm np}, i}^2(\mathcal T; \tau) \equiv \frac{1}{\mathcal T-\tau} \int_{0}^{\mathcal T-\tau}\left({\vec r}_i(t+\tau) - {\vec r}_i(t) \right)^2 dt \, ,
\end{equation}
with $\vec r_i (t)$ being the spatial position of the $i$-th nanoprobe at time $t$.
By tacitly assuming that the simulated trajectories are long enough such that the ``$\mathcal T\rightarrow\infty$'' limit is effectively reached,
the time average mean-square displacement is formally given by:
\begin{equation}
\Delta r_{{\rm np}, i}^2(\tau) \equiv \lim_{\mathcal T\rightarrow\infty} \Delta r_{{\rm np}, i}^2(\mathcal T; \tau) \, . \label{eq:npMSD-TimeAv} 
\end{equation}
The average over the ensemble of $N_{\rm np}$ nanoprobes is then given by:
\begin{equation}\label{eq:npMSD-EnsembleTimeAv}
\langle \Delta r_{\rm np}^2(\tau) \rangle \equiv \frac1{N_{\rm np}} \sum_{i = 1}^{N_{\rm np}} \, \Delta r_{{\rm np}, i}^2(\tau) \, . 
\end{equation}

\begin{figure*}
\centering{$\tau = 0.6\tau_{\rm MD}$}\\
\includegraphics[width=0.8\textwidth]{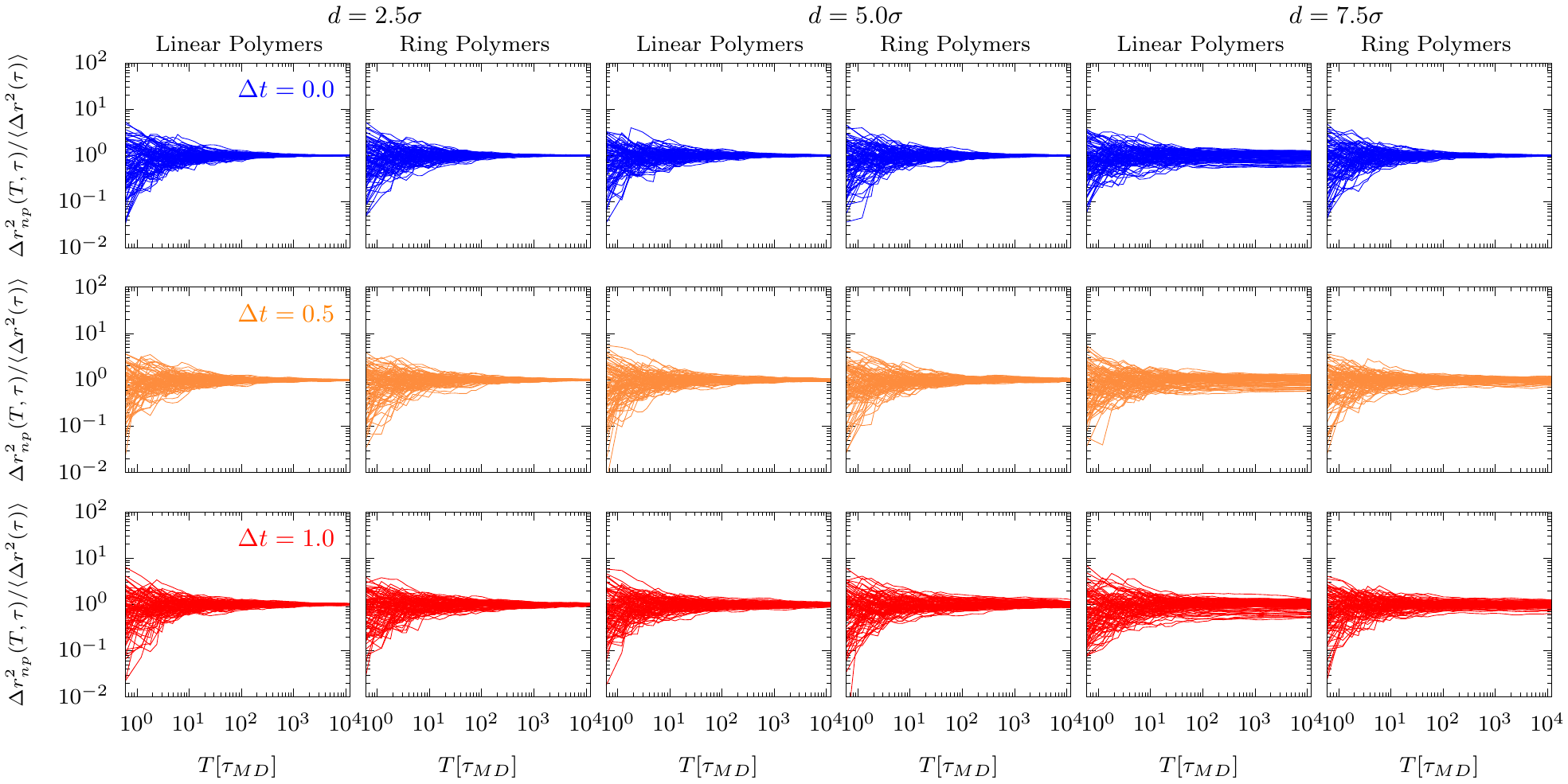} \\
\centering{$\tau = 6\cdot 10^2\tau_{\rm MD}$}\\
\includegraphics[width=0.8\textwidth]{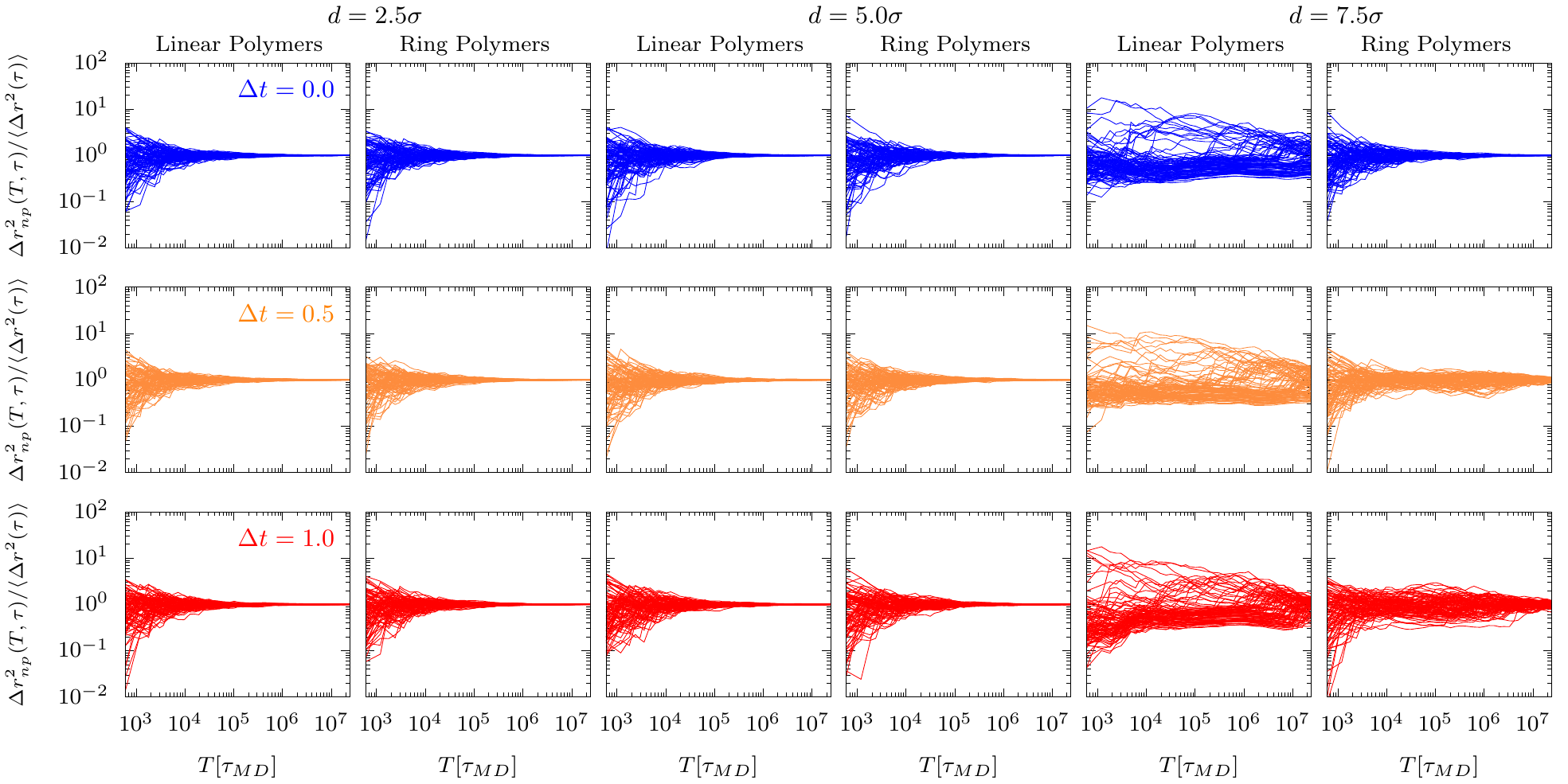}\\
\centering{$\tau = 6\cdot 10^3\tau_{\rm MD}$}\\
\includegraphics[width=0.8\textwidth]{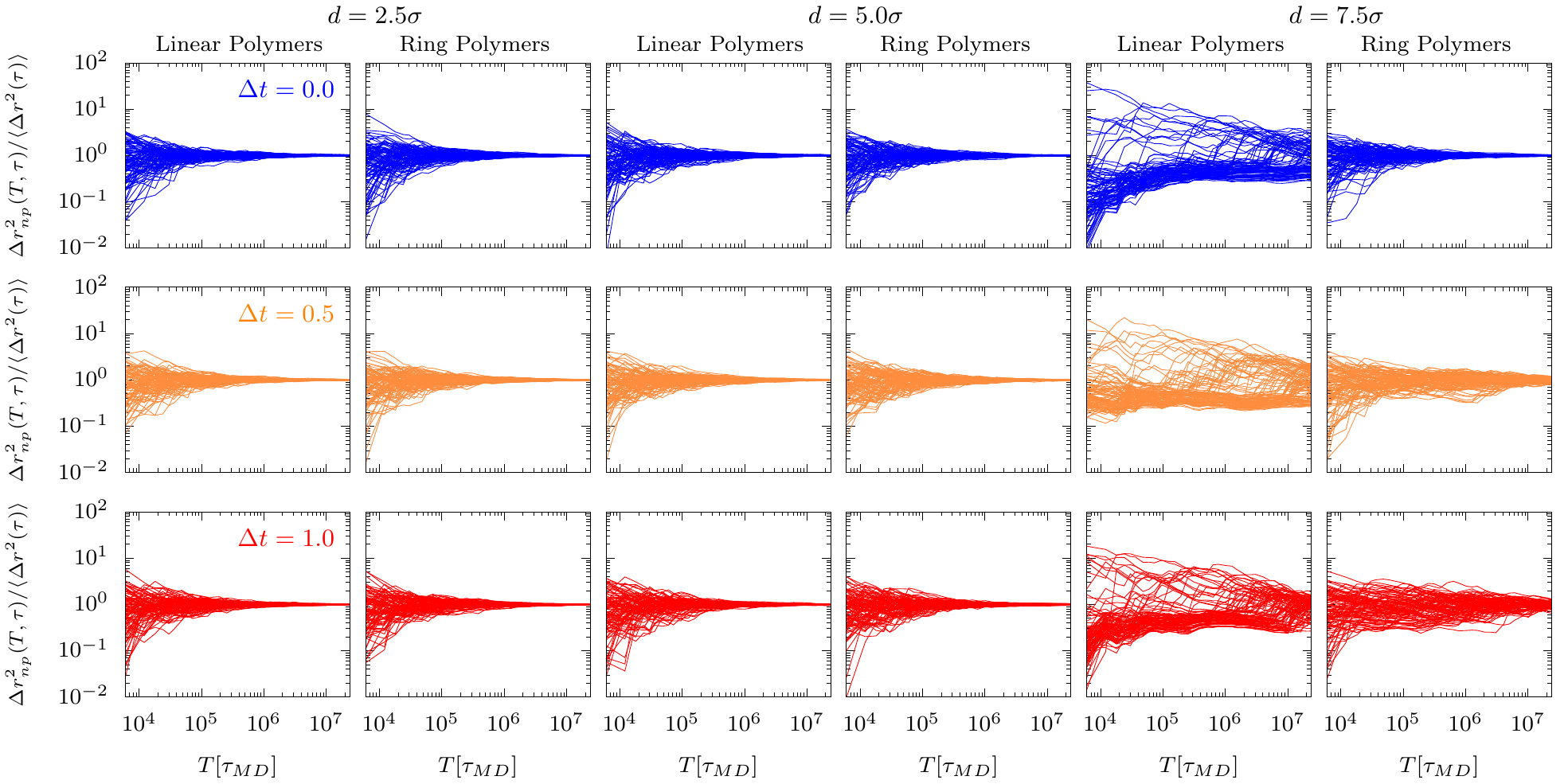} 
\caption{
\label{fig:SH}
Spatial heterogeneity, $\Delta r_{{\rm np}, i}^2({\mathcal T}; \tau) / \langle \Delta r^2 (\tau) \rangle$ (Eq.~\eqref{eq:SpatialHeterogeneity}), of nanoprobe mean-square displacements {\it vs.} the measurement time $\mathcal T$ for lag-times $\tau/\tau_{\rm MD} = 6\cdot10^{-1},10^2,10^3$ and nanoprobe diameters $d/\sigma=2.5,5.0,7.5$ (see legends). 
Each panel here contains $N_{\rm np}=100$ curves.
Color code is as in the rest of the paper.
}
\end{figure*}

In ergodic systems, Eq.~\eqref{eq:npMSD-TimeAv} should of course be independent from $i$.
This, however, might not be the case whenever dynamics is affected by long-range spatial correlations
as in glassy entangled polymer systems~\cite{MichielettoNahaliRosa2017,ChubakNatComm2020} or polymer nanocomposites~\cite{NahaliRosa2018,PapaleRosaPhysBiol2019}.
To detect such effects, we have measured the following ratios:
\begin{equation}\label{eq:SpatialHeterogeneity}
\Delta r_{{\rm np}, i}^2(\mathcal T; \tau) \, / \, \langle \Delta r_{\rm np}^2(\tau) \rangle \, \, \, \, \, \, (i=1, ..., N_{\rm np}) \, .
\end{equation}
Plots of the quantity Eq.~\eqref{eq:SpatialHeterogeneity} are shown in Fig.~\ref{fig:SH}.

Finally, motivated by the biased displacement orientation and following previous work~\cite{NahaliRosa2018,PapaleRosaPhysBiol2019},
we measure also the so called van-Hove~\cite{BerthierKobPRL2007} distribution function, $P(\tau; \Delta x)$, of the Cartesian components ($\alpha=x,y,z$) of nanoprobe spatial displacements for given lag-time $\tau$:
\begin{equation}\label{eq:DeltaxPDF}
P(\tau; \Delta x) \equiv \langle \delta[(r_\alpha(t+\tau) - r_\alpha(t)) - \Delta x] \rangle \, ,
\end{equation}
where $\delta$ is the Dirac's $\delta$-function.
For ordinary diffusion processes $P(\tau; \Delta x) = \frac{1}{\sqrt{2\pi \langle \Delta x^2 \rangle}} \exp\left( - \frac{\Delta x^2}{2\langle \Delta x^2 \rangle} \right)$ is Gaussian,
while correlated motion ({\it i.e.}, the one arising most typically in glassy and complex fluids~\cite{BerthierKobPRL2007,MichielettoNahaliRosa2017}) displays distributions with heavy tails.
Results for $P(\tau; \Delta x)$ are shown in Fig.~\ref{fig:DisplsPDF}.

\subsubsection{Single-chain dynamics}\label{sec:SingleChainDynamics}
\begin{figure*}
\includegraphics[width=0.9\textwidth]{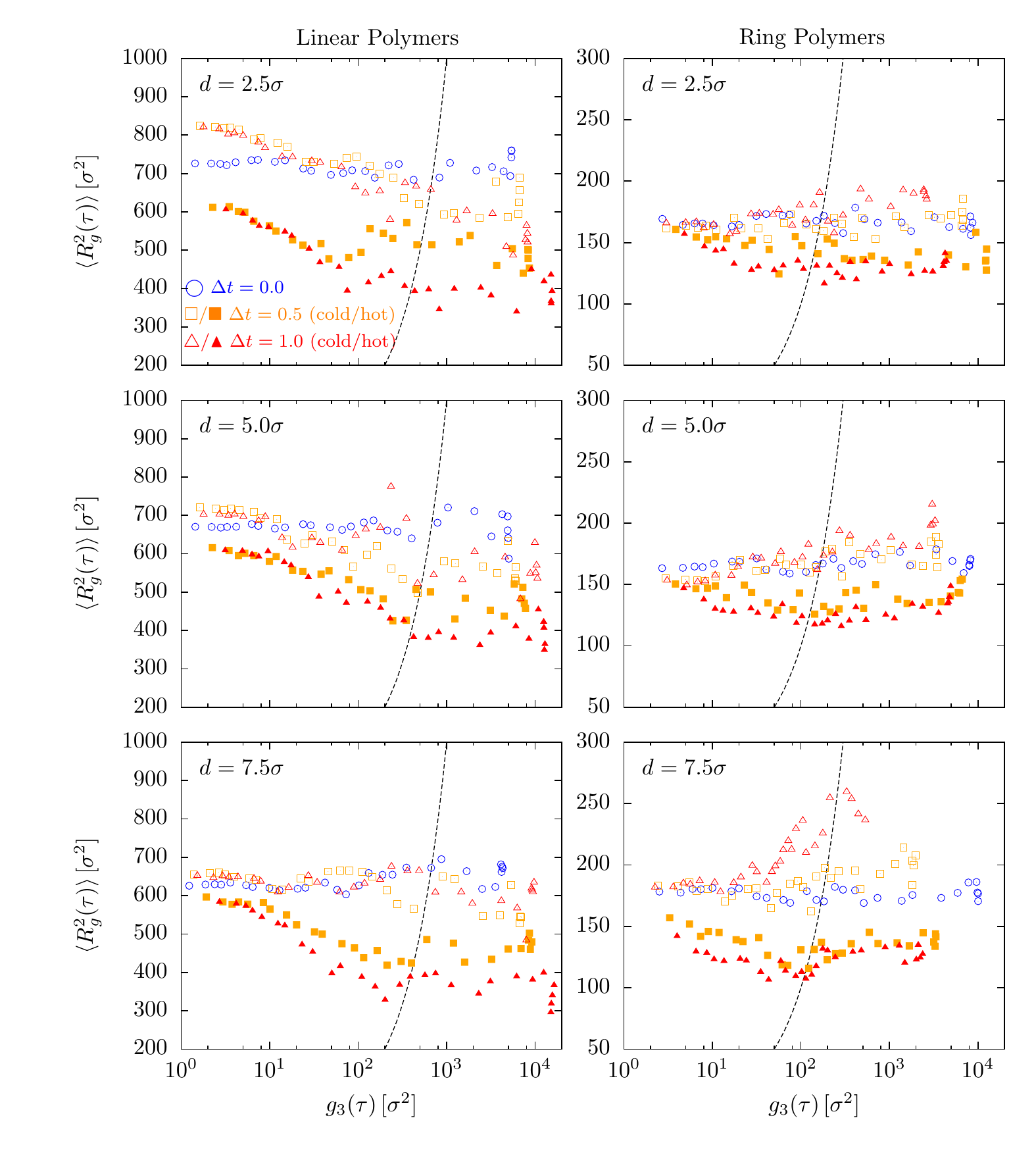}
\caption{\label{fig:msd_vs_Rg}
Parametric plot of the time evolution of the chain mean-square gyration radius, $\langle R_{g}^{2}(\tau) \rangle$ (average of Eq.~\eqref{eq:MeanSqGyrRad} on the ensemble of chains coupled to the same temperature $T$ for the single MD snapshot at time $\tau$), as a function of the mean-square displacement, $g_{3}(\tau)$ (Eq.~\eqref{eq:DefingeG3-EnsembleAv}), of the chain center of mass. 
The black dashed lines mark the positions where $g_{3} = \langle R_{g}^{2} \rangle$,
hence points to the right of the line demonstrate that the systems were run long enough to reach polymer displacements larger than the chain average gyration radius.
Color code is as in the main paper, with different colors corresponding to reduced temperatures $\Delta t = 0.0, 0.5, 1.0$.
Open/full symbols correspond to chains coupled to the cold/hot thermostat in passive/active mixtures (see legend).
}
\end{figure*}

\begin{figure*}
\includegraphics[width=0.485\textwidth]{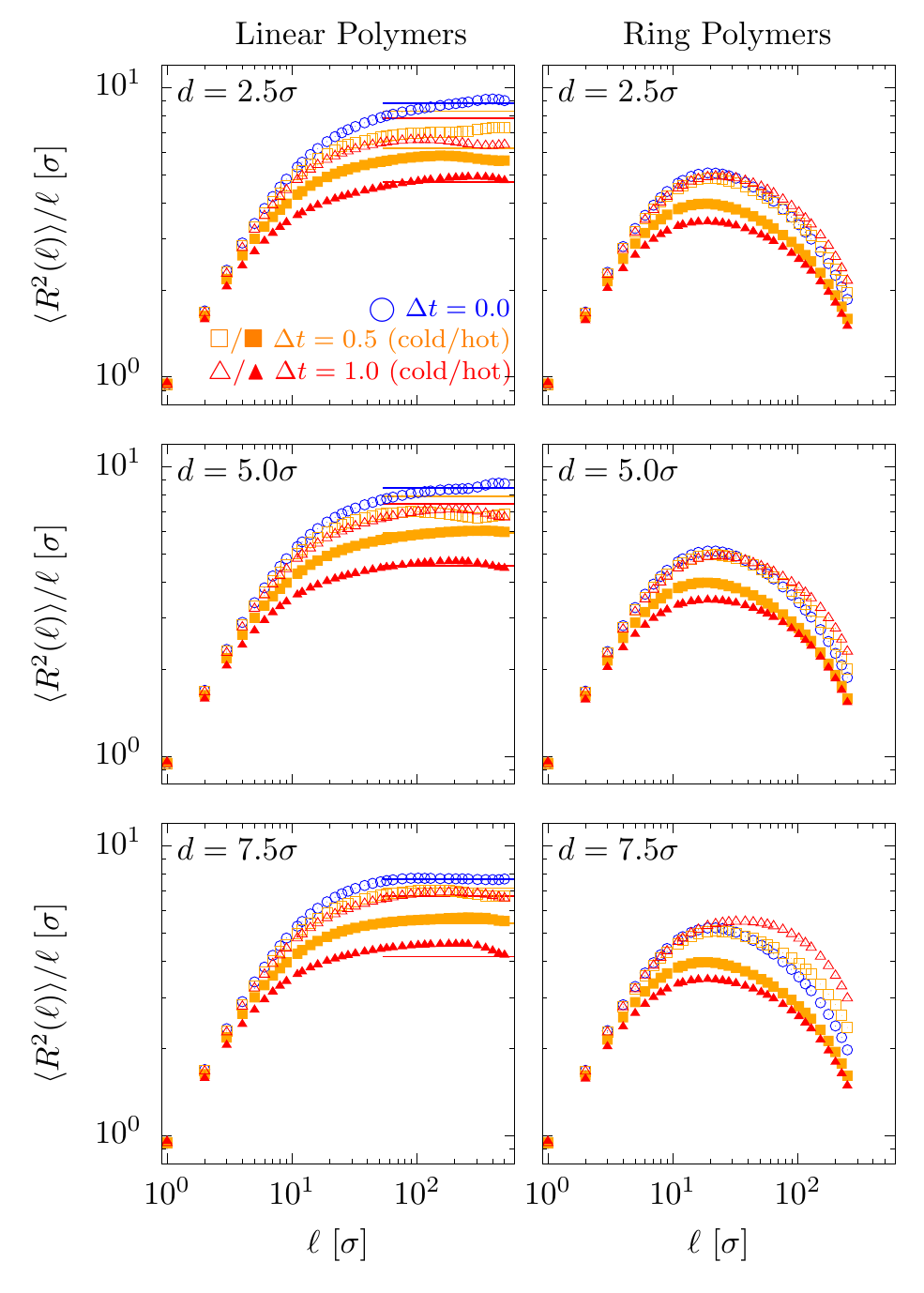}
\caption{\label{fig:R2vsN-Over-N}
$\langle R^2(\ell) \rangle / \ell$, mean-square end-to-end distances (Eq.~\eqref{eq:MeanSqEndToEnd}) of linear chains (l.h.s. panels) and rings (r.h.s. panels) normalized to corresponding monomer-monomer contour distances $\ell$.
Color code is as in the rest of the paper and choice of the symbols is as in Fig.~\ref{fig:msd_vs_Rg}.
For linear chains, the values of the plateaus at large $\ell$, $\ell_K \equiv \lim_{\ell\rightarrow\infty} \langle R^2(\ell) \rangle / \ell$, correspond to the Kuhn lengths of the respective chains~\cite{DoiEdwardsBook}:
the horizontal lines show results based on the formula $\ell_K(\langle T_{\rm ch} \rangle^{c, h}) \equiv \frac{\ell_K(\Delta t=0)}{\kappa_B \langle T_{\rm ch} \rangle^{c,h} / \epsilon}$,
where $\ell_K(\Delta t=0)$ comes from best fits of the passive-chain plateaus (on the interval $\ell/\sigma>100$) and $\langle T_{\rm ch} \rangle^{c, h}$ are the measured temperatures of cold/hot chains (see Table~\ref{tab:NpEffectiveTs}).
}
\end{figure*}

\begin{figure*}
\includegraphics[width=0.485\textwidth]{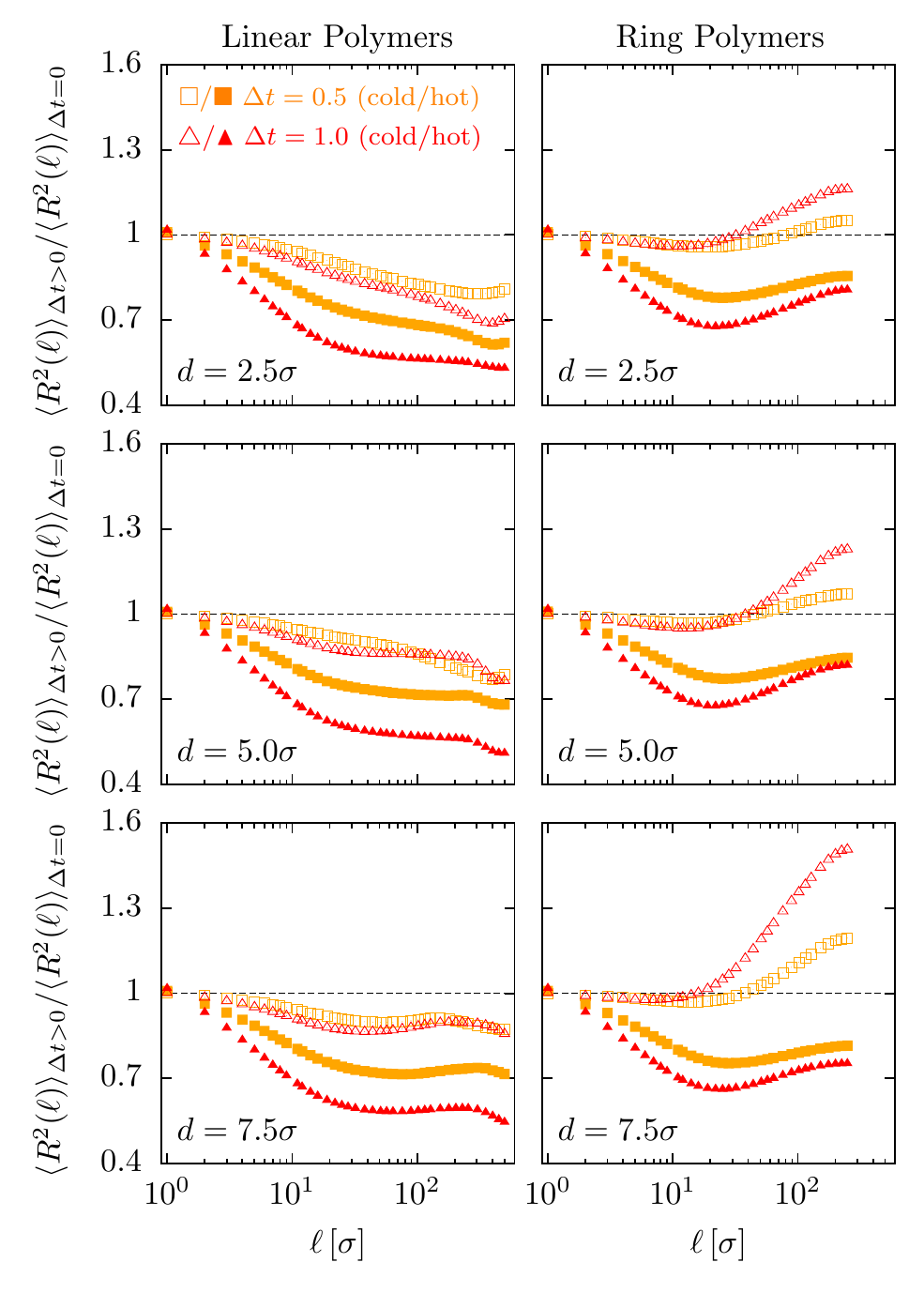}
\caption{\label{fig:R2vsN}
$\langle R^2(\ell) \rangle_{\Delta t >0} / \langle R^2(\ell) \rangle_{\Delta t = 0}$, mean-square end-to-end distances (Eq.~\eqref{eq:MeanSqEndToEnd}) as a function of the monomer-monomer contour distance $\ell$ for linear chains (l.h.s. panels) and rings (r.h.s. panels) in passive/active mixtures normalized with respect to the corresponding quantity measured in passive systems.
Color code is as in rest of the paper and choice of the symbols is as in Fig.~\ref{fig:msd_vs_Rg}.
}
\end{figure*}

Similarly to Eqs.~\eqref{eq:npMSD} and~\eqref{eq:npMSD-TimeAv}, we have considered the mean-square displacement, $g_{3, m}(\tau)$~\cite{DoiEdwardsBook,KremerGrestJCP1990}, of the centre of mass of the $m$-th chain in the solution:
\begin{equation}\label{eq:DefingeG3}
g_{3, m}(\tau) = \lim_{\mathcal T\rightarrow\infty} \frac{1}{\mathcal T-\tau} \int_{0}^{\mathcal T-\tau}\left({\vec r}_{{\rm cm}, m}(t+\tau) - {\vec r}_{{\rm cm}, m}(t) \right)^2 dt \, ,
\end{equation}
where ${\vec r}_{{\rm cm}, m}(t)$ is the coordinate of the centre of mass of the $m$-th chain.
As in static quantities (Sec.~\ref{sec:SingleChainStructure}), we take distinct averages of Eq.~\eqref{eq:DefingeG3} for the two polymer populations with the cold/hot thermostat (see Fig.~\ref{fig:msd_vs_Rg}):
\begin{equation}\label{eq:DefingeG3-EnsembleAv}
g_3^{\rm c,h}(\tau) = \frac1{M/2} \, \sum_{m=1}^{M/2}{\vphantom{\sum}}^{\rm c, h} \, g_{3, m}(\tau) \, .
\end{equation}
%


%
\begin{figure*}
\includegraphics[width=0.485\textwidth]{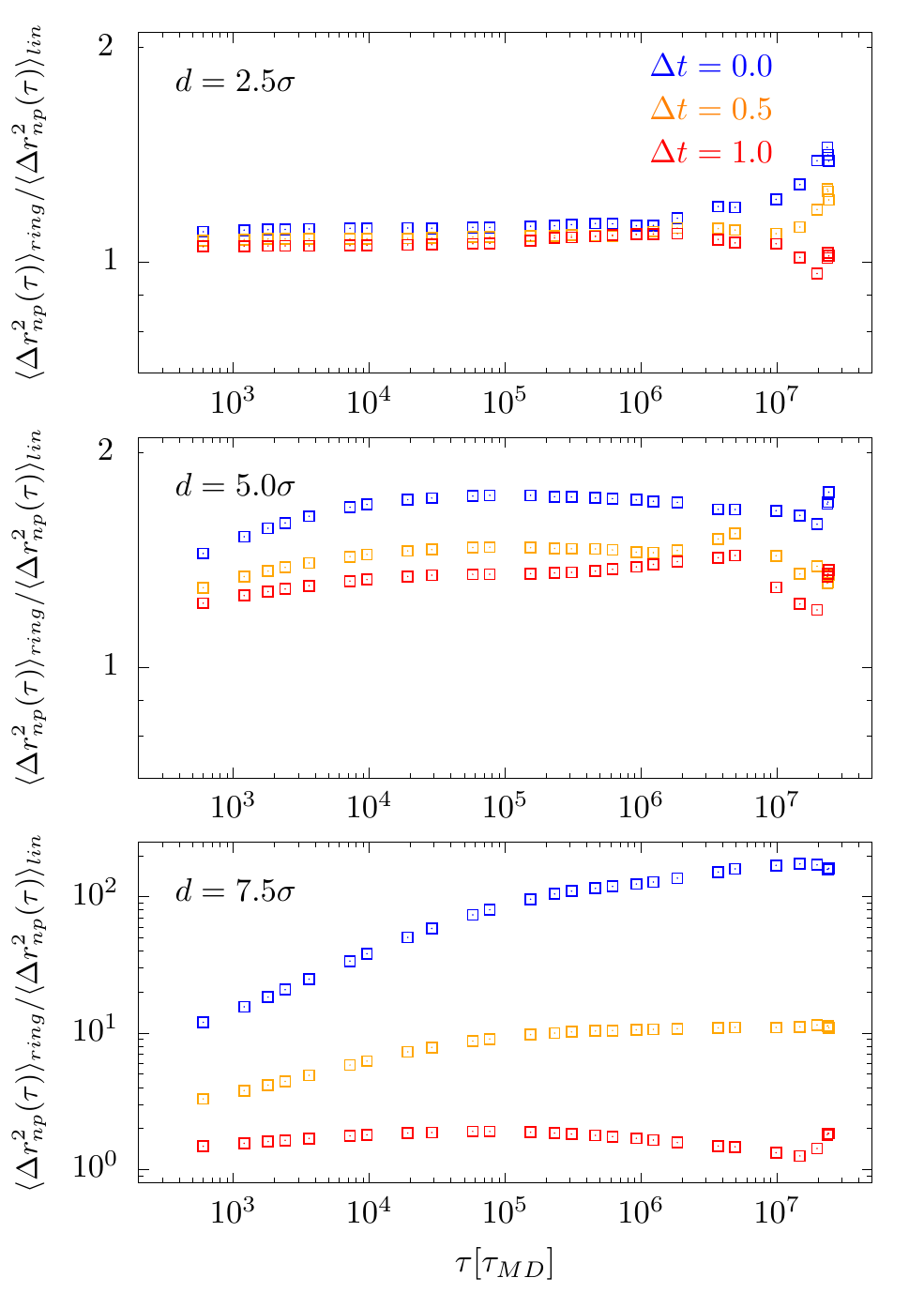} 
\caption{
\label{fig:npMSD-Ratios}
$\langle\Delta r_{\rm np }^2(\tau)\rangle_{\rm ring} / \langle\Delta r_{\rm np}^2(\tau)\rangle_{\rm lin}$, ratios of nanoprobe mean-square displacements (Eq.~\eqref{eq:npMSD-EnsembleTimeAv}) in rings {\it vs.} linear polymer solutions.
Results for increasing nanoprobe diameters $d$ (from top to bottom).
Color code is as in the rest of the paper.
Although diffusion in ring solutions is always larger than diffusion in linear solutions,
for $d=2.5\sigma$ and $d=5.0\sigma$ we notice a small yet clearly visible slow-down of the nanoprobes at increasing $\Delta t$.
Since the measured average temperatures of the nanoprobes are the same for the same $\Delta t$ ({\it i.e.}, they do not depend on polymer architecture, see Table~\ref{tab:NpEffectiveTs}),
we are tempted to ascribe this effect to the dependence of entanglements on chain flexibility~\cite{UchidaEveraersJCP2008,Svaneborg_Everaers_MAMOL20}.
In fact, in active-passive mixtures hot and cold chains of linear solutions are both more flexible than chains in fully passive counterparts (Figs.~\ref{fig:R2vsN-Over-N} and~\ref{fig:R2vsN}, l.h.s. panels) while in ring solutions (Fig.~\ref{fig:R2vsN-Over-N} and~\ref{fig:R2vsN}, r.h.s. panels) only hot chains bend more:
since more/less flexible chains are in general associated to less/more entangled polymers~\cite{UchidaEveraersJCP2008,Svaneborg_Everaers_MAMOL20} this may finally account~\cite{GeRubinsteinMacromolecules2017,RabinGrosbergNanoRheol2019} for the seen acceleration/deceleration of the nanoprobes.
On the other hand, this explanation sits on a definition of ``entanglements'' introduced and validated only for equilibrium system:
if it remains valid for out-of-equilibrium polymer solutions remains to be established, and more systematic investigations ought to be pursued in the future in this respect.
}
\end{figure*}
\begin{figure*}
\includegraphics[width=1.0\textwidth]{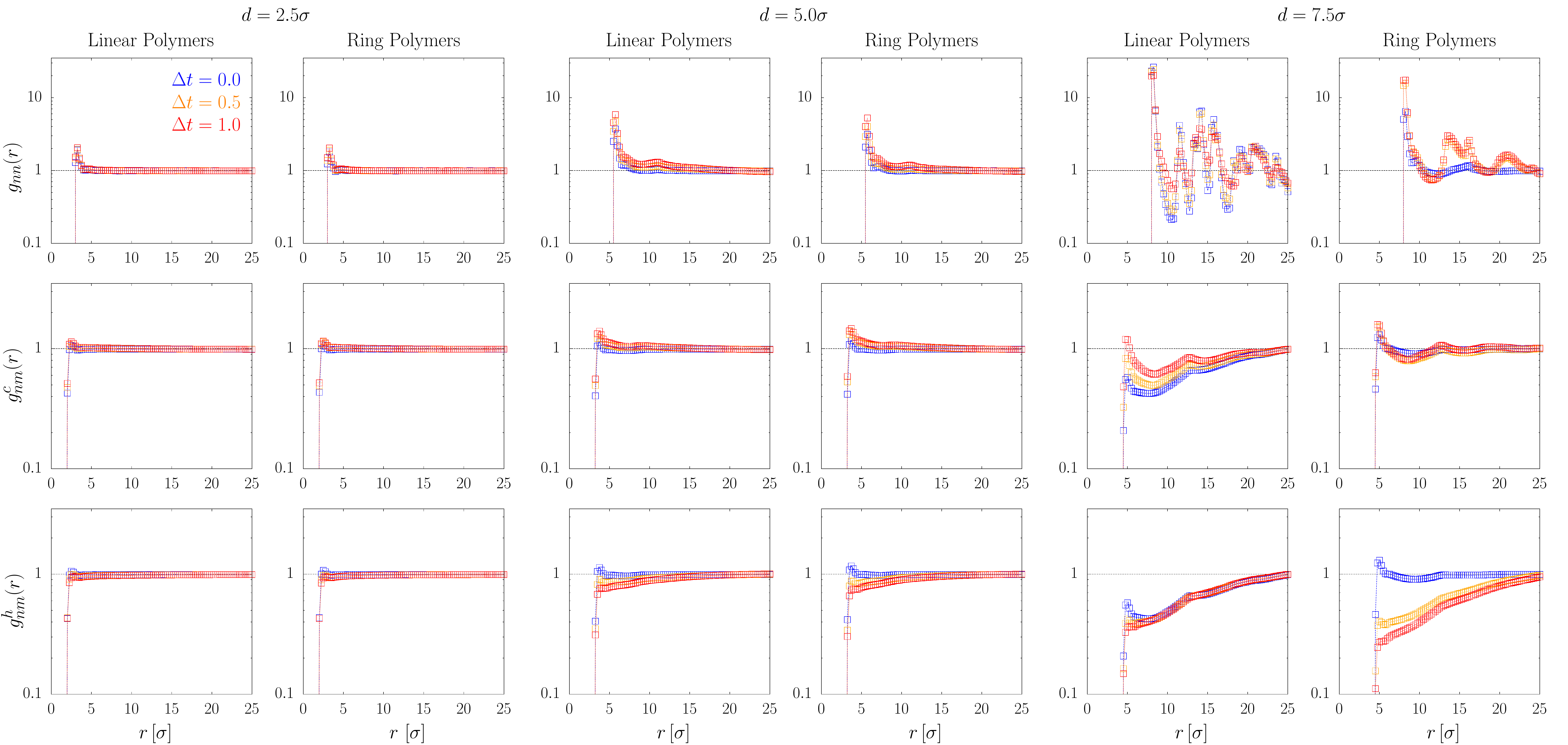}
\caption{
\label{fig:npGR}
Nanoprobe-nanoprobe ($g_{\rm nn}(r)$) and nanoprobe-monomer ($g_{\rm nm}^{\rm c}(r)$ and $g_{\rm nm}^{\rm h}(r)$) pair correlation functions
for nanoprobes of diameters $d/\sigma=2.5,5.0,7.5$ (see legends).
The superscripts indicate that the functions have been evaluated by separating the contributions of monomers coupled to the cold (c) or the hot (h) thermostat.
Color code is as in rest of the paper.
}
\end{figure*}
\begin{figure*}
\includegraphics[width=\textwidth]{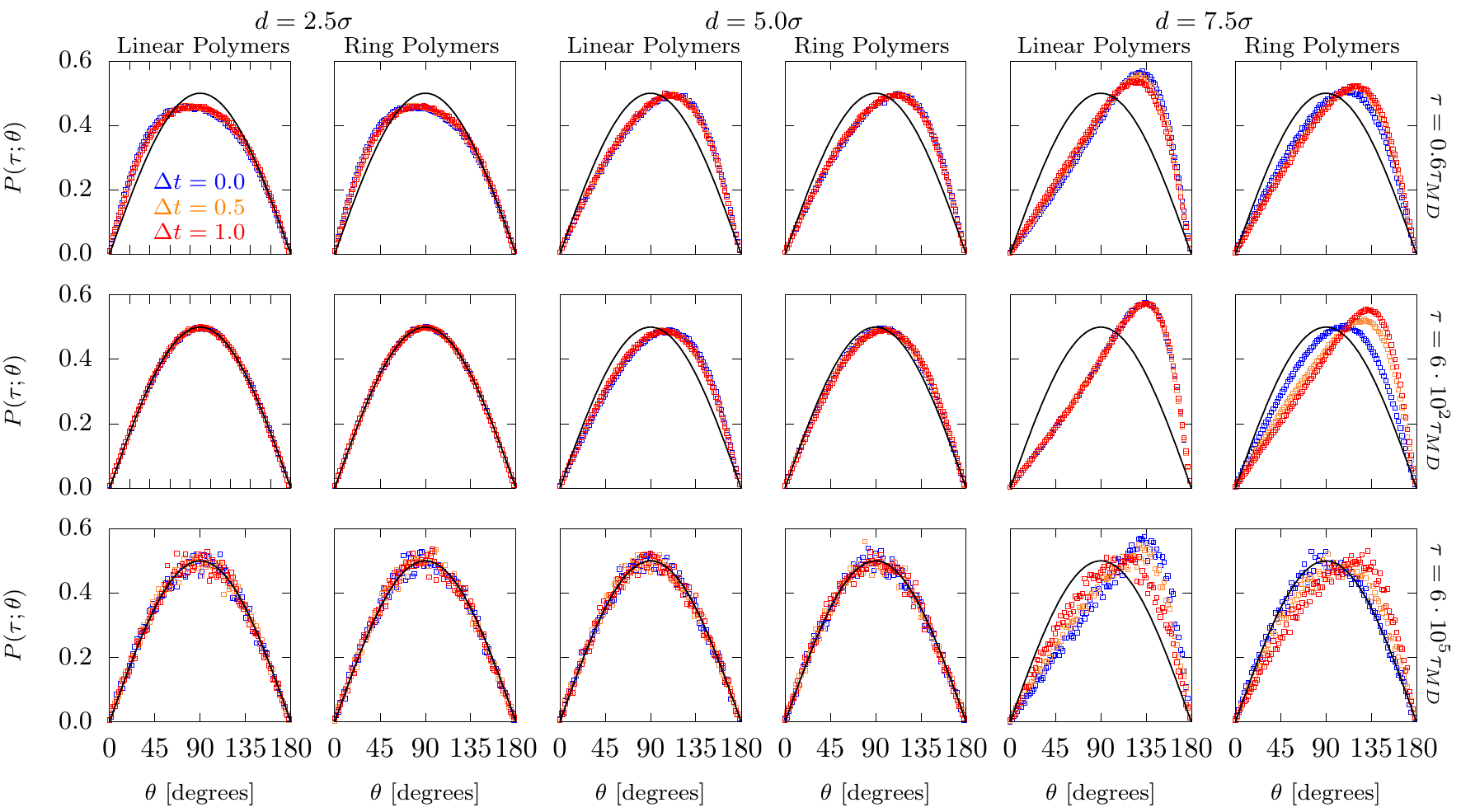} 
\caption{
\label{fig:pTheta_all}
Probability distribution functions, $P(\tau; \theta)$ (see Eq.~\eqref{eq:AngleCorrelFunct} in the main paper), of the angle $\theta$ between oriented spatial displacements of nanoprobes of diameters $d/\sigma=2.5,5.0,7.5$ and lag-times $\tau/\tau_{\rm MD}=6\cdot10^{-1},10^{2},10^{5}$ (see legends).
Color code is as in the rest of the paper. 
The black solid line is the function $P(\tau; \theta) = \frac12\sin\theta$ for randomly oriented vectors. 
}
\end{figure*}
\begin{figure*}
\includegraphics[width=\textwidth]{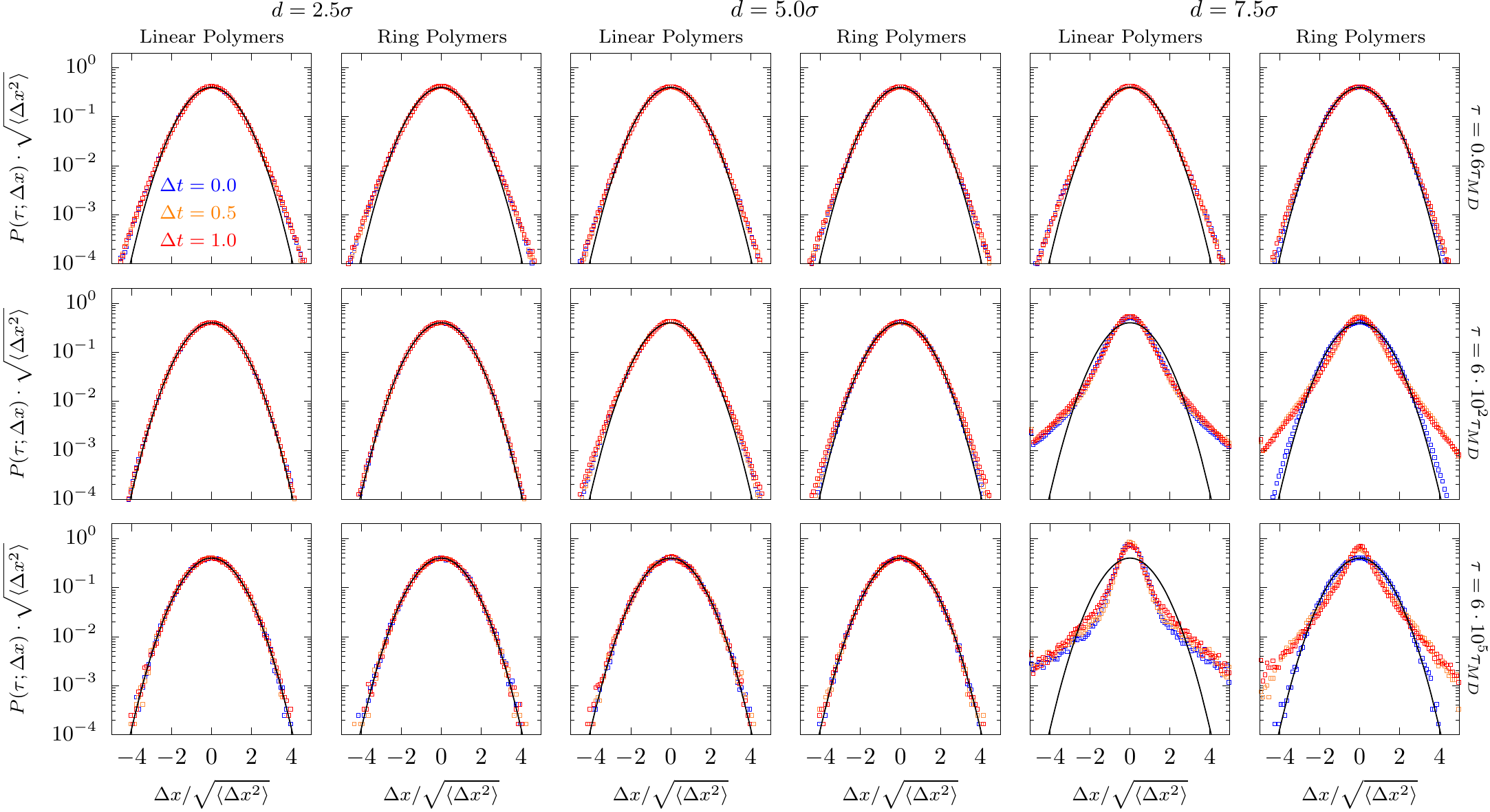} 
\caption{
\label{fig:DisplsPDF}
Probability distribution functions of one-dimensional nanoprobes displacements, $P(\tau; \Delta x)$ (Eq.~\eqref{eq:DeltaxPDF}), for the same representative lag-times $\tau$ 
as in Fig.~\ref{fig:pTheta_all} and nanoprobe diameters $d/\sigma=2.5,5.0,7.5$ (see legends).
Color code is as in the rest of the paper.
Black solid lines correspond to the theoretical Gaussian distribution function, $P(\Delta x)=\frac{1}{\sqrt{2\pi \langle \Delta x^2 \rangle}} \exp\left( - \frac{\Delta x^2}{2 \langle \Delta x^2 \rangle} \right)$, which is typical for ordinary diffusive processes. 
}
\end{figure*}
\end{document}